\documentclass[twocolumn]{aastex631}
\usepackage{newtxtext,newtxmath}
\usepackage[T1]{fontenc}
\usepackage{amsmath}
\usepackage[utf8]{inputenc}
\usepackage{graphicx}
\usepackage{float}
\usepackage[version=4]{mhchem}
\usepackage{multirow}
\usepackage{verbatim}

\newcommand*{\deltaccf}{$\Delta$CCF~}

\newcommand*{\kp}{$K_{\mathrm{p}}$}
\newcommand*{\vsys}{$V_{\mathrm{sys}}$~}
\newcommand*{\vbary}{$V_{\mathrm{bary}}$~}

\newcommand*{\vpx}{$V_{\mathrm{px}}$~}
\newcommand*{\vpxa}{$V_{\mathrm{px}}$}
\newcommand*{\kms}{$\mathrm{km\,s^{-1}}$~}
\newcommand*{\kmsa}{$\mathrm{km\,s^{-1}}$}
\newcommand*{\npx}{$N_{\mathrm{px}}$~}

\newcommand*{\npxa}{$N_{\mathrm{px}}$}
\newcommand*{\phimax}{$\phi_{\mathrm{max}}$}
\newcommand*{\deltav}{$\Delta v_{\mathrm{p}}$}
\newcommand*{\nin}{$N_{\mathrm{in}}$~}
\newcommand*{\logl}{log$(L)$~}
\newcommand*{\nout}{$N_{\mathrm{out}}$~}

\begin{document}

\title{Feasibility of High-Resolution Transmission Spectroscopy for Low-Velocity Exoplanets}

\correspondingauthor{Connor J. Cheverall / Nikku Madhusudhan}
\email{cjc218@ast.cam.ac.uk / nmadhu@ast.cam.ac.uk}

\author[0009-0005-4330-7197]{Connor J. Cheverall}
\affiliation{Institute of Astronomy, University of Cambridge, Madingley Road, Cambridge CB3 0HA, UK}

\author[0000-0002-4869-000X]{Nikku Madhusudhan} 
\affiliation{Institute of Astronomy, University of Cambridge, Madingley Road, Cambridge CB3 0HA, UK}

\begin{abstract}

In recent years, high-resolution transmission spectroscopy in the near-infrared has led to detections of prominent molecules in several giant exoplanets on close-in orbits. This approach has traditionally relied on the large Doppler shifts of the planetary spectral lines induced by the high velocities of the close-in planets, which were considered necessary for separating them from the quasi-static stellar and telluric lines. In this work we demonstrate the feasibility of high-resolution transmission spectroscopy for chemical detections in atmospheres of temperate low-mass exoplanets around M dwarfs  with low radial velocity variation during transit. We pursue this goal using model injection and recovery tests with H- and K- band high-resolution spectroscopy of the temperate sub-Neptune TOI-732~c, observed using the IGRINS spectrograph on Gemini-S. We show that planetary signals in transit may be recovered when the change in the planet's radial velocity is very small, down to sub-pixel velocities. This is possible due to the presence of the planetary signal in only a subset of the observed spectra. A sufficient number of out-of-transit spectra can create enough contrast between the planet signal and telluric/stellar contaminants that the planet signal does not constitute a principal component of the time-series spectra and can therefore be isolated using PCA-based detrending without relying on a significant Doppler shift. We additionally explore novel metrics for finding such signals, and investigate trends in their detectability. Our work extends the scope of high-resolution transmission spectroscopy and creates a pathway towards the characterisation of habitable sub-Neptune worlds with ground-based facilities.

\end{abstract}


\section{Introduction} \label{intro} 

High-resolution transmission spectroscopy has been at the forefront of the chemical characterization of exoplanetary atmospheres in recent years. The method's ability to resolve transmission spectra into individual spectral lines can break the degeneracy associated with broad and overlapping molecular bands, allowing for chemical detections with increased confidence. In this way, constraints on chemical compositions and atmospheric properties have been placed on a range of exoplanets \citep{snellen_orbital_2010, brogi_signature_2012, birkby_detection_2013, giacobbe_five_2021}.

Prominent CNO molecules exhibit strong absorption features in the near-infared (NIR) which can be inferred using high-resolution transmission spectra \citep{snellen_orbital_2010, brogi_signature_2012, birkby_detection_2013, alonso-floriano_multiple_2019, cabot_robustness_2019, guilluy_exoplanet_2019, giacobbe_five_2021, carleo_gaps_2022}. Atmospheric profiles, processes, and dynamics have also been constrained for these planets using this technique. There are several ground-based high-resolution spectrographs in use globally which span the NIR, including CARMENES \citep{quirrenbach_carmenes_2016, quirrenbach_carmenes_2018}, CRIRES(+) \citep{kaeufl_crires_2004, dorn_crires_2014, dorn2023}, GIANO \citep{oliva_giano-tng_2006, origlia_high_2014}, IGRINS \citep{yuk_preliminary_2010, park_design_2014}, HDS/Subaru \citep{noguchi_high_2002}, and SPIRou \citep{the_spirou_team_spirou_2018, donati_spirou_2020}.

When observing NIR spectra using high-resolution spectrographs on ground-based telescopes, any planet signal is buried in contaminating telluric and stellar features. A central challenge is therefore isolating the planet signal from such spectral contaminants, the features of which are significantly deeper. Commonly this detrending is done using principal component analysis (PCA) \citep{de_kok_detection_2013, brogi_retrieving_2019,giacobbe_five_2021, holmberg_first_2022, vansluijs2023, lafarga2023}, or a similar algorithm SYSREM \citep{tamuz_correcting_2005, mazeh_sys-rem_2007, birkby_discovery_2017, cabot_robustness_2019, spring_black_2022}. These algorithms have previously been successful for planets with a comparatively large change in radial velocity during the course of a transit. This enables the decoupling of the planet signal (which is Doppler-shifted as a function of orbital phase) from the telluric and stellar features (which remain relatively stationary in wavelength as the night progresses) \citep{birkby_exoplanet_2018}. Since a PCA-based detrending approach is expected to require the planet to undergo a sufficiently large change in radial velocity during the transit, it follows that the regime of potential targets may be limited to shorter period, warmer planets. Therefore, to date, high-resolution transmission spectroscopy has predominantly been used to characterize the atmospheres of hot Jupiters. The subtraction of a telluric model, which has been used commonly in the optical, e.g. using Molecfit \citep{smette_molecfit_2015}, is often not able to sufficiently isolate the planet signal in the NIR.

Recently, exoplanetary science has entered an exciting age of detecting and characterizing smaller, lower mass exoplanets. Using high-resolution transmission spectroscopy, studies have aimed to characterize the atmospheres of Neptune-like planets, such as GJ 1214 b and GJ 3470 b, using both real \citep{crossfield2011, deibert2019} and simulated data \citep{gandhi_seeing_2020, hood_prospects_2020, genest_effect_2022}. Work has also been done for super-Earths, e.g. 55 Cnc e and GJ 486 b \citep{esteves2017, jindal2020, deibert2021, ridden-harper}. However, whilst such work has probed the limits of small scale heights and transit depths, little work has so far been done in extending this method to longer period, temperate planets. Chemical inferences from high-resolution spectra have not yet been achieved for longer period planets, such as the sub-Neptunes TOI-732 c (also known as LTT 3780 c) \citep{nowak2020, cloutier2020} and TOI-270 d \citep{gunther2019}. Despite having comparable transit durations, such planets have longer periods and lower orbital velocities (observed via the radial velocity semi-amplitude \kp) than hot Jupiters, and consequently they undergo a relatively small change in radial velocity over the course of a transit. This change is generally considered insufficient for a PCA-based detrending method to be successful in decoupling the planet signal from the wavelength-stationary telluric background. The goal of this paper is to examine this assumption by investigating the potential for using ground-based high-resolution transmission spectroscopy for the characterization of longer period, low-radial-velocity-change planets, which in this work we will often refer to as low-velocity planets for ease.

Successful detrending of ground-based high-resolution transmission spectra of low-radial-velocity-change exoplanets would offer various advantages and opportunities for future atmospheric characterization of smaller, temperate planets. 
Firstly, whilst transmission spectra can be observed using space-based telescopes, 
broad-band molecular features can be overlapping and degenerate. On the other hand, high-resolution ground-based spectroscopy with adequate signal-to-noise can resolve individual spectral lines, allowing for more confident identification of such molecular features.
Secondly, low-resolution spectra obtained from space
can be devoid of molecular features due to the damping effect of clouds and hazes in the planet's atmosphere \citep{sing2016, barstow2017, pinhas2019}. Whilst this is the case for a diverse range of planets, it is particularly true for cool sub-Neptune planets \citep{desert2011, bean2011, kreidberg2014}, where clouds/hazes limit our capacity to constrain their atmospheric compositions. However, at the higher resolutions possible from the ground, it is possible to probe spectral lines originating above the clouds and hazes \citep{gandhi_seeing_2020, hood_prospects_2020}. Whilst the James Webb Space Telescope (JWST) offers much higher resolution (up to R $\sim$ 3,000) \citep{ferruit2020} than the Hubble Space Telescope (HST) WFC3 and therefore may provide a promising window into characterizing the atmospheres of these sub-Neptune planets, ground-based high-resolution spectrographs offer significantly higher resolutions (R $\gtrsim$ 25,000) \citep{brogi_rotation_2016} and are therefore suited to probing cool and cloudy planets. 

Finally, planets orbiting M-dwarf stars are the ideal targets for atmospheric characterization studies of small, low-mass exoplanets, as the reduced stellar size allows for a more pronounced planetary transmission spectrum. Low-mass planets are also found to be very common around M-dwarf stars. The challenge with the study of M-dwarf systems is their high stellar activity, which in the low-resolution regime may be degenerate with planetary features \citep{rackham2023}. However, it has been found that this degeneracy can be broken at higher resolutions, such that confident chemical inferences can be made \citep{genest_effect_2022}. High-resolution spectroscopic studies of temperate planets around these active stars would therefore offer significant advantages over space-based, low-resolution observations.

This paper aims to extend the scope of high-resolution transmission spectroscopy in the NIR to a new regime of exoplanets. As a case study, we analyse high-resolution transmission spectra of TOI-732 c, a sub-Neptune planet orbiting an M-dwarf star, obtained using IGRINS. This planet is a Hycean candidate \citep{madhusudhan2021} orbiting an M-dwarf host star. To date, no chemical detections have been made using ground-based high-resolution spectroscopy in the atmospheres of sub-Neptune planets, or for any planet around an M-dwarf star. The paper is organized as follows. In Section \ref{methods} we describe standard methods used for analysing NIR high-resolution transmission spectra. In Section \ref{low_velocity_regime} we show that these methods can successfully be extended to the low-planetary-velocity regime in particular cases, and introduce a new additional metric which may be able to distinguish planet signals of certain molecules from correlated telluric, stellar, or noisy contaminants. In Section \ref{likelihood_section} we compare our findings using cross-correlation spectroscopy to those obtained when using a Bayesian likelihood approach. We investigate the trends and limitations of high-resolution spectroscopy for temperate planets in Section \ref{limits}, and identify the key observational and physical parameters driving our ability to successfully correct for tellurics using PCA. We then discuss what our results mean in the context of partial transits and emission spectroscopy in Section \ref{partial_emission}. Our results are summarised in Section \ref{summary}.

\section{General Methods} \label{methods}

In this section we describe the general methods by which a chemical detection can be obtained using high-resolution transmission spectroscopy in the NIR. We first examine the quality of the spectral observations used here. After cleaning and normalization, the spectra are then corrected for tellurics using principal component analysis, before being cross-correlated with a high-resolution model planet spectrum in order to determine a detection S/N for each model. Each of these steps is outlined in more detail in this section. 

\subsection{Observations} \label{observations}

\begin{table}
\centering
\begin{tabular}{lc}
\hline \hline
 Parameter & Value\\ \hline
 \hline
 $P$ & $12.252131^{+0.000072}_{-0.000064}$ d\\
 $T_{0}$ & $2458546.8492^{+0.0016}_{-0.0017}$ BJD\\
 $R_{\mathrm{star}}$ & $0.382\pm0.012$ $\mathrm{R_{\odot}}$ \\
 $R_{\mathrm{p}}$ & $2.42\pm0.10$ $\mathrm{R_{\oplus}}$ \\
 \vsys & $0.1954 \pm 0.0005$ \kms\\ 
 $a$ & $0.0762\pm0.0034$ au \\
 $i$ & $89.08^{+0.11}_{-0.13}$ $^{\circ}$ \\
 $e$ & 0.115$^{+0.07}_{-0.065}$ \\
 $T_{14}$ & $1.79\pm0.21$ h \\
 \kp & $70 \pm 20$ \kms \\ 
 \hline
\end{tabular} 
\caption{System properties of TOI-732 c. Values adopted from \citet{nowak2020}, except \vsys and \kp~which are from \citet{cloutier2020}. We note that \citet{cloutier2020} gives a value for $T_{14}$ of $1.392^{+0.049}_{-0.050}$ h, but here we use the more conservative value from \citet{nowak2020}.}
\label{toi732c_parameters}
\end{table}

For our present exploratory study we consider simulated planet spectra injected into high-resolution time-series observations of a known sub-Neptune, TOI-732 c (LTT 3780 c). The observed spectra were obtained using the IGRINS spectrograph at Gemini-S as part of GO Program GS-2021A-Q-201 (PI: D. Valencia) (Cabot et al. 2024, in press). We consider the observations of one transit obtained on the night of February 23 2021. TOI-732 c was discovered by the Transiting Exoplanet Survey Satellite (TESS) mission \citep{Ricker2015} and followed up with ground-based radial velocity observations with CARMENES and HARPS \citep{nowak2020, cloutier2020}. The system parameters are shown in Table \ref{toi732c_parameters}. For the transit duration there are multiple values reported in the literature: 1.4 h \citep{cloutier2020} and 1.8 h \citep{nowak2020}. We adopt the latter value of 1.8 h to be conservative. There have been no atmospheric characterisation studies of this planet to date, although several recent studies have considered simulated studies with JWST \citep{madhusudhan2021,constantinou_characterising_2022}. 

IGRINS is a high-resolution spectrograph mounted on the 8\,m Gemini South telescope at Cerro Pachón, Chile \citep{yuk_preliminary_2010, park_design_2014}. IGRINS observes the NIR spectral range, covering $\sim$1.45 - 2.45 $\mu$m, split into 54 spectral orders, and with a resolution of $R \sim$ 45,000. The present time-series observations include 33 spectra (A/B pairs) obtained over 2.8 hours. This includes $21\pm2$ spectra during the transit, calculated using the transit duration given in Table \ref{toi732c_parameters}. The different value for $T_{14}$ in \cite{cloutier2020} would instead suggest that  $15_{-0}^{+2}$ of the spectra are in transit. 
The airmass varied between 1.05 and 1.18, and the magnitude of the barycentric velocity correction varied between 3.2\,\kms and 2.8\,\kmsa, over the course of the observations. 

\subsection{Data cleaning and Normalization} \label{clean_norm}
We follow the methods in \cite{cheverall_robustness} to clean the spectra of bad pixels and outliers, with each spectral order treated independently. We begin with the reduced spectra of the observations made available through the Gemini/IGRINS archive. These are the time-series spectra divided by the spectrum of an A0V telluric standard star obtained using the same setting. We remove 24 orders (0-4, 24-37, and 49-53 inclusive) due to poor quality data or lack of a well-defined continuum. This leaves 30 spectral orders for the remainder of the analysis. To normalize the spectra, we fit a second order polynomial to the continuum of each spectral order and exposure. However, due to the relatively low temperature of the M-dwarf host star, stellar molecular features appear in the spectral continuum which in some cases are difficult to remove with a fitted polynomial. Despite this, with the exception of \ce{H2O}, most of the molecules typical of temperate planets are not expected to be present in the stellar photosphere, and therefore any uncorrected stellar spectral features should not interfere with our injection and recovery tests, the results of which we compare with and without injections. A more suitable normalization function would be desirable for real chemical inferences however.

\subsection{Telluric Correction} \label{detrending}

For ground-based observations of spectra in the NIR, the planet transmission signal is embedded in contaminating telluric and stellar features as well as correlated noise from other sources, such as the instrument. In order to isolate the planet signal from such contaminants, which are significantly stronger, we detrend our spectra using principal component analysis (PCA), as is commonly done in previous works \citep{de_kok_detection_2013, giacobbe_five_2021, lafarga2023}. This number of principal components subtracted, or `PCA iterations applied', is calculated by optimizing the S/N from a noise subtracted, differential CCF, \deltaccf \citep{holmberg_first_2022, spring_black_2022}, in order to avoid the introduction of an optimisation bias towards the signal for which we are searching \citep{cheverall_robustness}. This is as opposed to optimizing for the detection S/N directly, which has the potential to introduce bias \citep{cabot_robustness_2019, spring_black_2022, cheverall_robustness}. We calculate the optimum PCA iteration for all of the 3 models considered here, and find optimal values of 3, 5, and 7 for \ce{NH3}, \ce{CH4}, and \ce{H2O}, respectively. 

\begin{figure}
\includegraphics[width=0.5\textwidth]{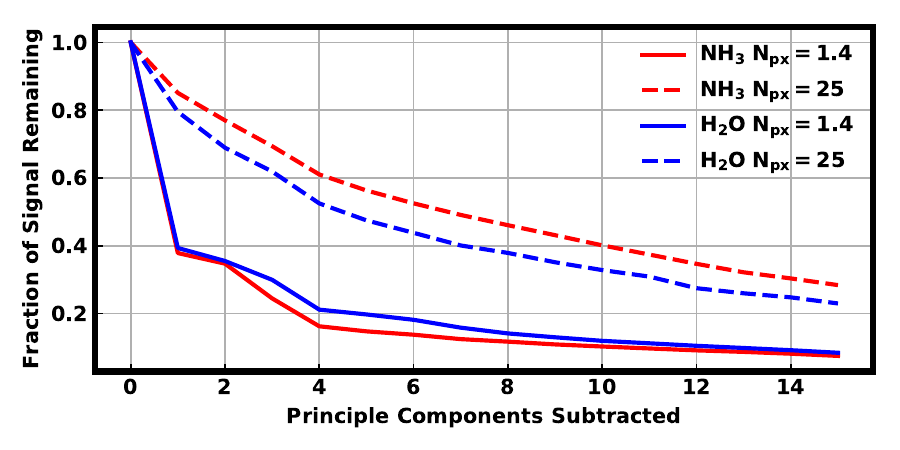}
\caption{Effect of PCA detrending on the planet signal for different planet orbital velocities and molecules. Here we compare the degradation of \ce{NH3} and \ce{H2O} signals by PCA for a temperate planet such as TOI-732 c (solid) with that for a planet with a higher orbital velocity, e.g., that typical of a hot Jupiter such as HD 189733 b (dashed). The signal is eroded faster for the planet with lower radial velocity change over the transit, as would be expected. The planetary pixel shift, \npx, introduced later in Section \ref{low_velocity_regime}, is the number of instrument pixels by which the planet signal is Doppler-shifted during the transit. Note that the absolute signal strength of \ce{NH3} is stronger than that for \ce{H2O}, with all cases here rescaled with respect to their own initial strength.}
\label{fig:signal_degredation}
\end{figure}

The planetary signal, which is quasi-stationary in wavelength,  constitutes a lower frequency signal in the time-series spectra than a typical high-velocity planet signal which is shifting across the instrument pixels throughout the observing night. We find that the degradation of the planet signal is therefore significantly faster at lower velocities (Figure \ref{fig:signal_degredation}). It is hence reasonable to expect that fewer principal components may need to be subtracted in order to maximally recover low-velocity planet signals, since the subtraction of too many components will remove the planet signal alongside the quasi-stationary telluric and stellar lines.

\subsection{High-resolution model spectra} \label{modelling}

To compute model templates for the transmission spectra of the sub-Neptune TOI-732 c considered in this study, we follow a similar approach to that of \citet{cheverall_robustness}, which we briefly reiterate here. The transmission spectrum is modelled using the AURA atmospheric modelling code for exoplanets \citep{pinhas_retrieval_2018}. The model computes line-by-line radiative transfer in transmission geometry assuming a plane-parallel H$_2$-rich atmosphere in hydrostatic equilibrium, over a pressure range of $10^{-7}$ - $100$ bar. The chemical composition and temperature structure are free parameters in the model. We generate the spectra considering one molecule at a time, assuming no clouds/hazes and an isothermal temperature profile. We consider a nominal temperature of 300 K, close to the equilibrium temperature of TOI-732 c \citep{madhusudhan2021}. The spectra are computed at high resolution (\textit{R} $\gtrsim$ 10$^5$) in the near-infrared spanning the IGRINS spectral range (1.4-2.5 $\mu$m). 

We consider opacity contributions due to prominent molecules expected in temperate H$_2$-rich atmospheres (H$_2$O, CH$_4$, and NH$_3$) and assuming a nominal mixing ratio of 10$^{-4}$ for each molecule. The molecular cross-sections were obtained following the methods of \cite{gandhi_genesis_2017} using absorption line lists from the following sources: H$_2$O \citep{barber2006,rothman_hitemp_2010}, CH$_4$ \citep{yurchenko_exomol_2014}, and NH$_3$ \citep{yurchenko_variationally_2011}. We also include collision-induced absorption from H$_2$-H$_2$ and H$_2$-He \citep{borysow1988,orton2007,abel2011,richard_new_2012}.
As with the observed spectra, the model spectra are separated into orders and then normalised using a polynomial fit. This is then followed by convolution with the point spread function of the instrument. Whilst the absolute depths of absorption lines are lost after normalisation, their relative depths and positions are conserved.

\subsection{Signal extraction using cross-correlation} \label{signal_extraction_cc}

After detrending the spectra of unwanted telluric and stellar features, as described in Section \ref{detrending}, we are left with residuals containing the remaining planet signal. Each individual line in this planetary spectrum has S/N $<<$ 1, so information from each needs to be combined in order for the signal to be characterized. Therefore, the residuals are typically cross-correlated with a Doppler-shifted model, shifted as a function of planetary velocity, to give a cross-correlation function CCF as a function of radial velocity and orbital phase. This planetary radial velocity can be given by:
\begin{equation} \label{v_p}
V_{\mathrm{p}} = K_{\mathrm{p}} \sin (2\pi\phi) + V_{\mathrm{sys}} - V_{\mathrm{bary}} + V_{\mathrm{offset}}
\end{equation}
where \kp is the semi-amplitude of the planet's orbital motion, \vsys is the systemic velocity of the planetary system, \vbary is the barycentric velocity correction where a positive value indicates the Earth is moving towards the star, and $V_{\mathrm{offset}}$ accounts for any radial velocity offset, originating from sources such as planetary atmospheric winds. 

The standard approach to extract a signal from the CCF is to calculate a detection S/N for each point in \kp - \vsys space. 
The signal is calculated by summing the CCF values along a trail defined by a given point in \kp - \vsys space, and then divided by a noise estimate obtained from the standard deviation of CCF values away from this trail, as described in \cite{cheverall_robustness}. A strong cross-correlation signal at the known orbital parameters of the planetary system would indicate a detection. The left-hand panel of Figure \ref{fig:gamma_sn1} shows the resulting S/N maps when the detrended residuals are cross-correlated with \ce{NH3}, \ce{CH4}, and \ce{H2O} models. There are no significant (S/N > 3) cross-correlation signals recovered at the planetary velocities at which we later inject model planet spectra (marked by crosshairs) to create synthetic data sets in Section \ref{low_velocity_regime}, with only a tentative feature seen at \vsys $\sim20$\,\kms for \ce{NH3}.

These are the basic techniques with which we complete injection and recovery tests in Section \ref{low_velocity_regime}. This is then repeated using more in-depth Bayesian analysis of the spectra in Section \ref{likelihood_section}. 

\section{Signal recovery in the low-velocity regime} \label{low_velocity_regime}

In this section we consider the recovery of planetary signals in the low-velocity, long-period regime. As a case study, we inject model planetary spectra into the observations of TOI-732 c described in Section \ref{observations}. We show that with sufficient out-of-transit spectra, it is indeed possible to recover these signals for certain molecules, and hence detect chemicals in the atmospheres of low-velocity planets, using the techniques outlined in Section \ref{methods}. We highlight the challenges and limitations inherent to this regime of exoplanet, and introduce a new metric to aid in the identification and verification of cross-correlation signals.

\subsection{Simulation of spectra} \label{injection}

In order to investigate the limits of high-resolution transmission spectroscopy for long-period planets we simulate the spectra of such targets. Previous works using simulations of high-resolution spectra have often involved the generation of a synthetic data set, built from models of the component parts \citep{hood_prospects_2020, gandhi_seeing_2020, genest_effect_2022}. In this work, rather than generate a synthetic data set, we use the real observations described in Section \ref{observations} and inject a high-resolution model planetary signal. This has the benefit of ensuring we inject synthetic planetary spectra into data sets containing realistic and comprehensive sources of noise. It does however limit our control and knowledge over the noise in the data. 

Along with the telluric absorption and correlated noise from the instrument, the broad-band stellar features present in the spectral observations of an M-dwarf star, as discussed in Section \ref{clean_norm}, also modulate the planetary spectral lines. To simulate this modulation we initially inject our planetary spectra prior to normalization.

When simulating a data set, we must select which of the spectra will be in-transit, i.e. the spectra into which we inject a model planet spectrum, and which will be out-of-transit. This selection is initially made symmetrically in phase as follows. For $N$ in-transit and $M$ out-of-transit spectra, the $N$ spectra with the $N$ smallest values of $|\phi|$ become the in-transit spectra, whilst the next smallest $M$ exposures become the out-of-transit spectra for use in detrending. Any remaining exposures (in the case where $N$ + $M$ < 33) are removed from the data set in this instance. 
Note that our phases are near-symmetrical about zero, with the central exposure near mid-transit, meaning that when an odd number of in-transit spectra are included in our simulated data set, there are an equal number of out-of-transit spectra on each side of the transit. Partial transits are investigated in Section \ref{partial_emission}.

\subsection{Initial recovery of signals at low velocities}

We now complete injection and recovery tests with synthetic atmospheric model signals of \ce{NH3}, \ce{CH4}, and \ce{H2O}, which represent a range of telluric contributions. These models are individually injected, following the steps in Section \ref{injection}, with the expected planetary parameters given in Table \ref{toi732c_parameters} (\kp~of 70\,\kmsa, with 21 in-transit and 12 out-of-transit spectra). We aim to recover these injected signals as described in Section \ref{methods}, subtracting 3, 5, and 7 principal components, respectively, to detrend in accordance with the \deltaccf metric.

\begin{figure*}
    \centering
    \includegraphics[width=0.49\textwidth]{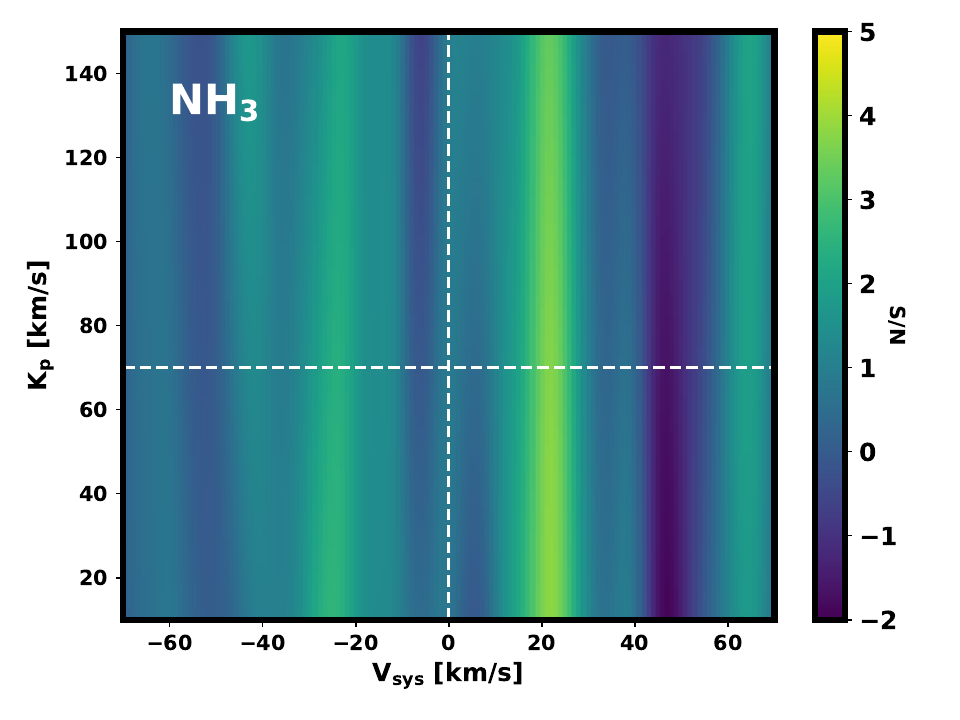}
    \includegraphics[width=0.49\textwidth]{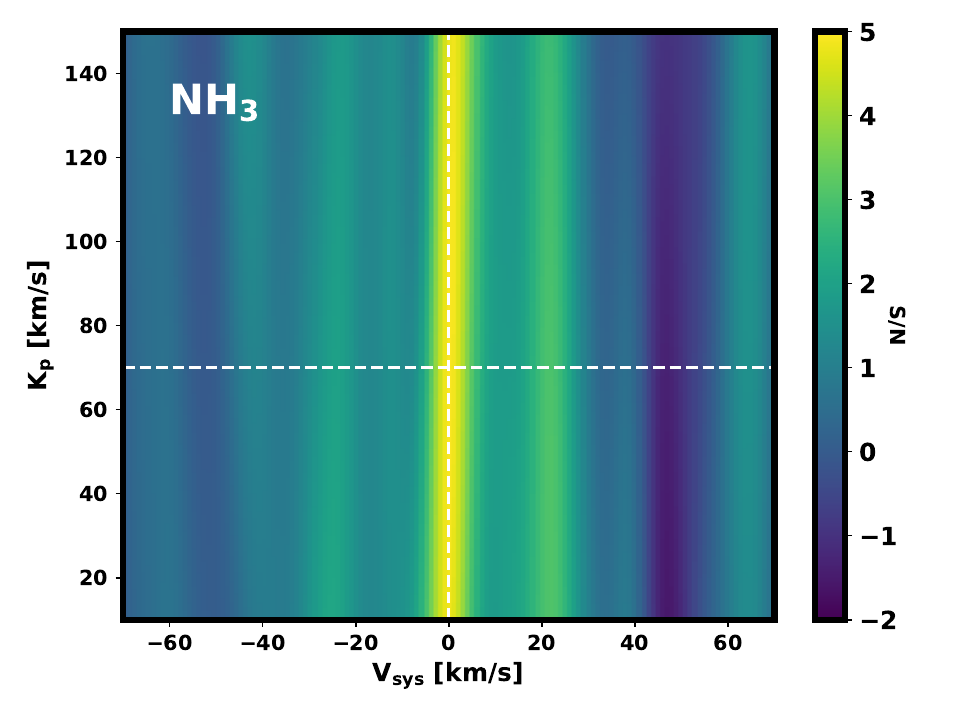}
    \includegraphics[width=0.49\textwidth]{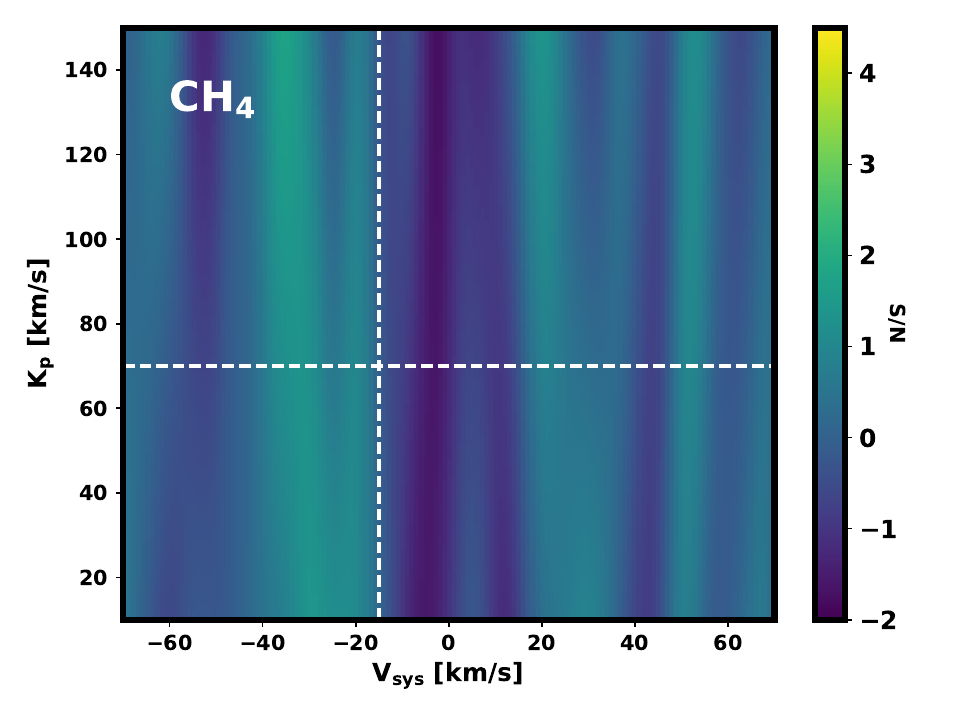}
    \includegraphics[width=0.49\textwidth]{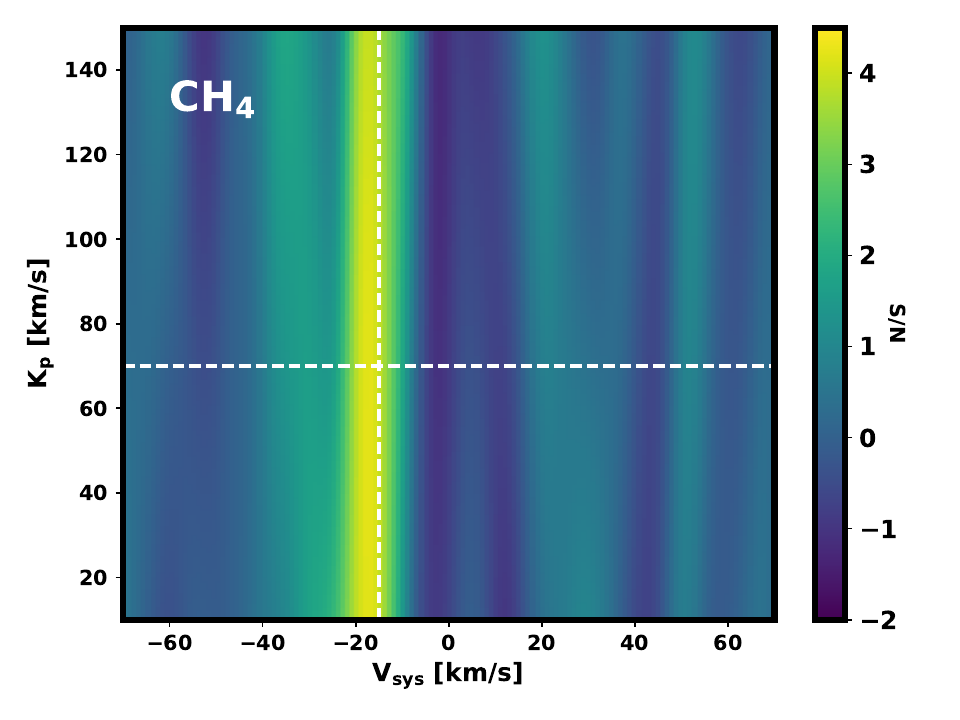}
    \includegraphics[width=0.49\textwidth]{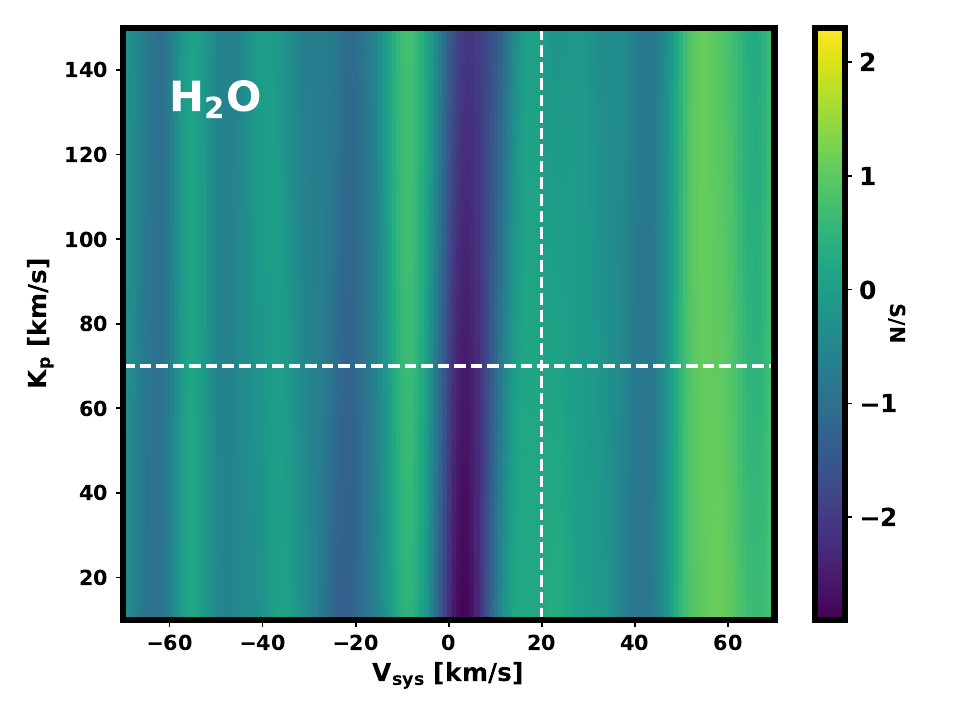}
    \includegraphics[width=0.49\textwidth]{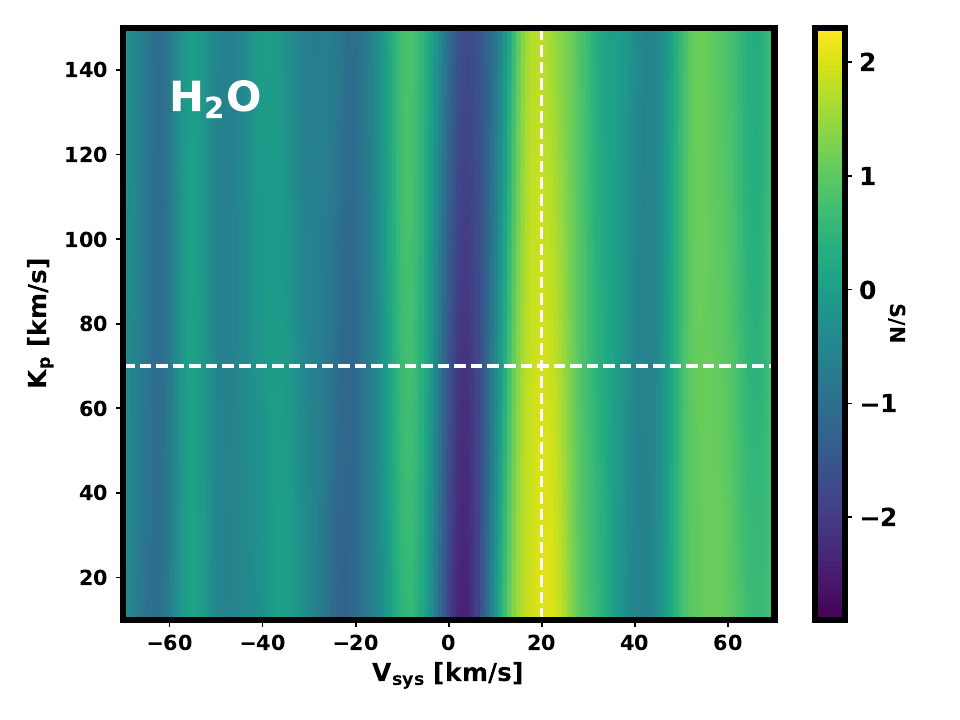}
    \caption{Successful recovery of low-velocity planet signals with varying S/N for each molecule. Shown is the S/N across \kp - \vsys space when cross-correlating the detrended residuals, first without (left) and then with (right) prior atmospheric model injection, with the relevant atmospheric models. 3, 5, and 7 principal components were removed in detrending for each of \ce{NH3}, \ce{CH4}, and \ce{H2O}, respectively, in accordance with the \deltaccf metric. Left hand panel: without prior injection, no significant signals are seen where we later inject model planetary spectra (marked by crosshairs) to create synthetic data sets, with only a tentative feature seen at \vsys $\sim20$\,\kms for \ce{NH3}. Right hand panel: the injected signals are recovered, with varying strengths for each molecule, at the correct \vsys, with S/N values of 5.0, 4.2, and 2.1 for \ce{NH3}, \ce{CH4}, and \ce{H2O}, respectively. These signals are unconstrained in \kp. In each case, the signal has been injected into 21 in-transit spectra with a planetary velocity of \kp~= 70\,\kmsa.}
    \label{fig:gamma_sn1}
\end{figure*}

The cross-correlation results are shown in Figure \ref{fig:gamma_sn1}, with detection significances of 5.0, 4.2, and 2.1 found for \ce{NH3}, \ce{CH4}, and \ce{H2O}, respectively. The cross-correlation signals shown do not coincide with any of the positive background noise seen without injection; since the only difference between these two cases is the injection of a planetary signal prior to detrending, it is likely that the signals seen here are indeed recoveries of those injections. Similar injection and recovery tests are completed for the high velocity case used in Figure \ref{fig:signal_degredation}, with results shown for comparison in Figure \ref{fig:high_velocity}.

We find that we are able to recover these injections independently of the systematic velocity. This is demonstrated in Figure \ref{fig:gamma_sn1}, where the value of \vsys at which each model is injected and recovered is varied for completeness, and the recovery of \ce{NH3} at a \vsys of 0\,\kms eliminates a large \vsys as responsible for the separation of planet and telluric spectra. In order to test the robustness of the recovered signals, we verify that the detection S/N is somewhat stable against the number of principal components removed in detrending.

We additionally repeat this using the raw spectra without the A0V correction used here (Figure \ref{fig:raw}), and find that we continue to recover the low-velocity planet signals with S/N values consistent with before (5.0 for \ce{NH3}, 3.4 for \ce{CH4}, and 1.6 for \ce{H2O}). We also note that similar results are found when the signals are injected after normalization.

Our inability here to recover a significant (S/N > 3) \ce{H2O} signal may perhaps be due to the weaker spectral features of this molecule compared to \ce{NH3} and \ce{CH4}. The signal strengths shown in Figure \ref{fig:signal_degredation} are normalised by their initial values at 0 PCA iterations, with the initial \ce{H2O} signal significantly weaker than that for \ce{NH3}. The reduced detection significance for this molecule is also seen when simulating a high-velocity injection (Figure \ref{fig:high_velocity}), and no significant increase in S/N is observed when the signal is injected and recovered at artificially high systematic velocities, e.g., >300 \kms, to decouple it from telluric features. The \ce{H2O} signal with features amplified by 2.5x is detected with S/N > 4 in Figure \ref{fig:gamma_sn2}.

\begin{figure}
    \centering
    \includegraphics[width=0.5\textwidth]{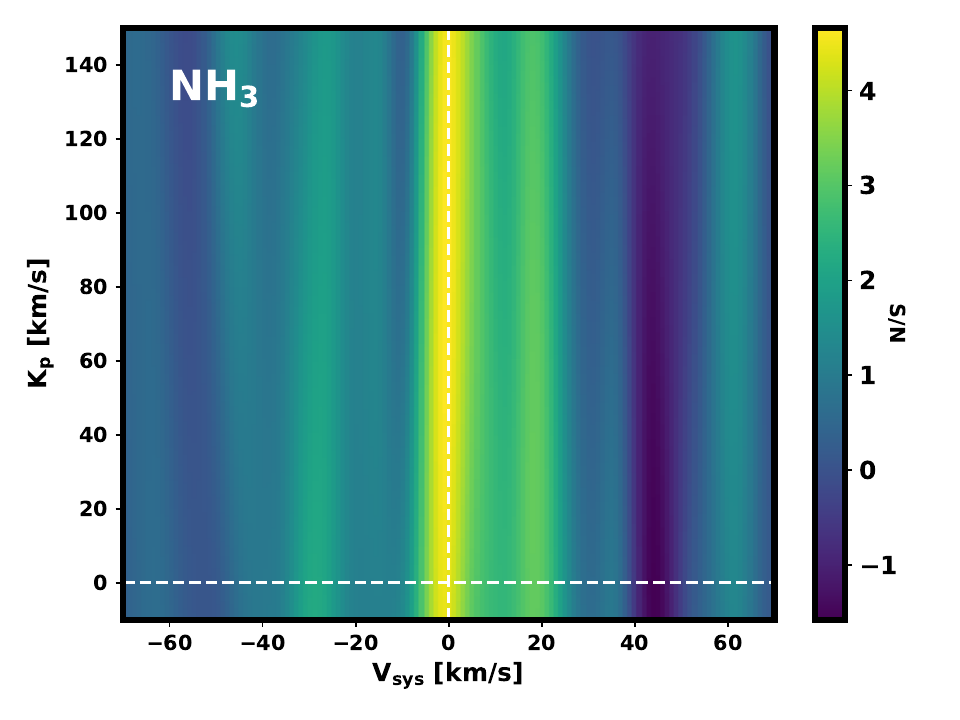}
    \caption{Recovery of an \ce{NH3} signal injected with zero orbital velocity. The detection S/N across \kp - \vsys space is shown when an \ce{NH3} signal is injected into the spectra with no semi-amplitude radial velocity (\kp = 0\,\kms) and no barycentric velocity correction and then recovered as in Figure \ref{fig:gamma_sn1}. A signal with a S/N of 4.7, unconstrained in \kp, is recovered at the correct \vsys.}
    \label{fig:gamma_sn0}
\end{figure}

The relatively small phase range and radial velocity semi-amplitude of TOI-732 c means that the radial velocity change during the transit is very small, less than the velocity resolution of the instrument (see Table \ref{rv_change}). However, we have shown that we are still able to recover injected signals of varying strengths for this planet using PCA-based detrending. We now probe the limits of low-radial-velocity-change planets, by injecting a model of \ce{NH3} into the spectra at \kp~= 0\,\kms while simulating zero barycentric velocity correction, and once again attempting to recover. After again subtracting 3 principal components in detrending, a signal with S/N of 4.7 is found (Figure \ref{fig:gamma_sn0}), despite there being no change in the radial velocity of the planet during the transit, i.e. the planet signal undergoes no wavelength shift relative to the telluric contaminants.

We have shown that little to no relative change in the Doppler-shift of the planet signal compared to that of the telluric spectrum does not prohibit PCA-based detrending. 
Therefore, it is possible to chemically characterize the atmospheres of temperate, longer period planets, and the assumption that the target requires a distinguishable change in radial velocity during the transit is not necessarily true. However, despite being able to recover the signals, we are unable to constrain any information about the \kp~of the planet in any of the cases shown in Figures \ref{fig:gamma_sn1} or \ref{fig:gamma_sn0}.

\subsection{Uncertainty in radial velocity semi-amplitude} \label{uncertainty}

For the methods described in Section \ref{signal_extraction_cc}, in order to determine the \kp~of a cross-correlation signal we must measure the change in radial velocity of the planet during the course of the transit, \deltav. This measurement is converted to \kp, assuming an accurate value for the maximum phase during transit, \phimax, for comparison with the known orbital parameters of the planetary system. \phimax~is calculated from the transit duration $T_{14}$ and the period $P$. We are however limited by the resolution of the spectrograph instrument, with the width of each wavelength channel corresponding to a given velocity shift which we will henceforth refer to as the instrument pixel velocity, \vpxa. This propagates an intrinsic error into measurements of \deltav~and hence \kp. The number of instrument wavelength channels by which the planetary signal is Doppler-shifted during the transit, which we define as the planet pixel shift, \npxa, is given by:
\begin{equation} \label{npx_equation}
N_{\mathrm{px}} = \frac{\Delta v_{\mathrm{p}}}{V_{\mathrm{px}}} = \frac{2K_{\mathrm{p}}\sin(2\pi\phi_{\mathrm{max}}) - \Delta v_{\mathrm{bary}}}{V_{\mathrm{px}}}
\end{equation}
\begin{table}
\centering
\begin{tabular}{lcccc}
\hline \hline
Planet & \kp~/ \kms & \phimax & \deltav / \kms & Ref \\ \hline
\hline
HD 189733 b & 152.5 & 0.017 & 33 & B16;A10;A19 \\
HD 209458 b & 145 & 0.018 & 33 &  G21;K07;A12 \\
WASP-76 b & 196.5 & 0.044 & 107 & E20 \\
\hline
TOI-270 d & 70 & 0.004 & 3.5 & G19;VE21 \\
TOI-732 c* & 70 & 0.003 & 2.6 & N20;C20 \\
\hline
\end{tabular}
\vspace{5mm}
\caption{Typical changes in the radial velocity of planets during their transits. For reference, a typical instrument pixel velocity \vpx is around 1-2 \kms for resolutions of $R\sim50,000-100,000$. We here compare the hot Jupiters studied in \citet{cheverall_robustness} (HD 189733 b, HD 209458 b, WASP-76 b) to example sub-Neptunes (TOI-270 d and TOI-832 c). *indicates planet studied in this work. Note that contributions from the changing barycentric velocity correction are not considered here, and errors are not given as we are simply indicating the approximate pixel shifts for different regimes of planet. References: B16 \citep{brogi_rotation_2016}; A10 \citep{agol_climate_2010}; A19 \citep{addison_minerva-australis_2019}; G21 \citep{giacobbe_five_2021}; K07 \citep{knutson_using_2007}; A12 \citep{albrecht_obliquities_2012}; E20 \citep{ehrenreich_nightside_2020}; G19 \citep{gunther2019}; VE21 \citep{eylen2021}; N20 \citep{nowak2020}; C20 \citep{cloutier2020}.}
\label{rv_change}
\end{table}

The radial velocity change of example exoplanets during their transits is shown in Table \ref{rv_change}. 
The IGRINS ($R \sim 45,000$) data we use here has an instrument pixel velocity of \vpx $\approx 2$ \kmsa. Typical hot Jupiters, such as those given in Table \ref{rv_change}, may therefore have planetary pixel shifts \npx of $\sim 10-100$ for this instrument. Other high-resolution spectrographs with resolutions $R \sim 50,000 - 100,000$ may increase this pixel shift for a given planet. For example, CARMENES ($R \sim 80,000$) observations of WASP-76 b \citep{landman_detection_2021, sanchez-lopez_searching_2022} correspond to a planetary pixel shift of \npx $\sim 80$, whilst CARMENES observations of HD 189733 b \citep{alonso-floriano_multiple_2019} correspond to \npx $\sim 25$. We use a simulated \npx of 25 for the high-velocity comparisons in Figures \ref{fig:signal_degredation} and \ref{fig:high_velocity} (corresponding to \deltav~$\approx$ 50 \kms for this instrument).
In such cases one can obtain a constrained determination of the \kp~of a planetary signal, although the finite instrument resolution is significant enough that prominent tails on signals in \kp~space remain (see Figure \ref{fig:high_velocity}). However, for sub-Neptune planets such as TOI-732 c, the change in radial velocity during the transit is typically much smaller. This is mainly driven by significantly smaller values for \phimax~due to the increased period of the planets. For this planet and instrument, accounting for the decrease in \vbary over the transit, the planet pixel shift is \npx $\sim$ 1.4\,px (consisting of 1.3\,px from the planet's orbit (Table \ref{toi732c_parameters}) and 0.1\,px from the barycentric velocity variation). We can therefore no longer constrain \kp~and so any measurement of \kp~is now redundant. %
We note that we are considering the number of wavelength channels crossed by the planet signal during transit, rather than the number of resolution elements. This is larger than the latter, typically by a factor of around 3, meaning it may be difficult to resolve the velocity shifts for planets with \npx $\lesssim$ 3, such as TOI-732 c.

The inability to constrain \kp~could be an issue when aiming to confirm the robustness of a detection. Having only one useful parameter, \vsys, may not be sufficient to distinguish between spurious signals and a robust detection of a species in the atmosphere of the planet. For example, telluric and stellar features, and correlated noise from other sources, will likely produce similar unconstrained cross-correlation signals in \kp - \vsys space. To help us identify genuine planet cross-correlation signals from spurious noise, it would therefore be useful to include another parameter in our analysis.

\subsection{Constraining transit duration in the low-velocity regime} \label{novel}
As described in the previous section, it can be difficult to differentiate cross-correlation signals arising from low-velocity planets from those attributed to residual quasi-static telluric and stellar signals. This is because we can no longer separate the planetary signal from these contaminants based on a well constrained and large value for \kp~characteristic of hot Jupiters. However, for planets whose change in radial velocity during the transit is small, it may still be possible to verify the origin of cross-correlation signals using a different metric. Whereas the planetary signal has a fixed transit duration, spurious signals in the CCF from e.g. residual telluric and stellar features, may not exhibit the same phase-dependence from being constrained to the transit. 

To test this, we now treat \phimax~as a free parameter and explore over \phimax~- \vsys space for constant \kp. All spectra are included in detrending, but for each point in \phimax~space the in-transit spectra are defined individually when summing the CCF in time. A S/N is then calculated for this set of in-transit spectra for each \vsys, eventually giving S/N as a function of \phimax~- \vsys space, or \nin-~\vsys space where \nin is the number of in-transit spectra. This is done for constant \kp, here set to the known \kp~of the planet in question (70 \kmsa, see Table \ref{toi732c_parameters}).
\begin{figure}
    \centering
    \includegraphics[width=0.5\textwidth]{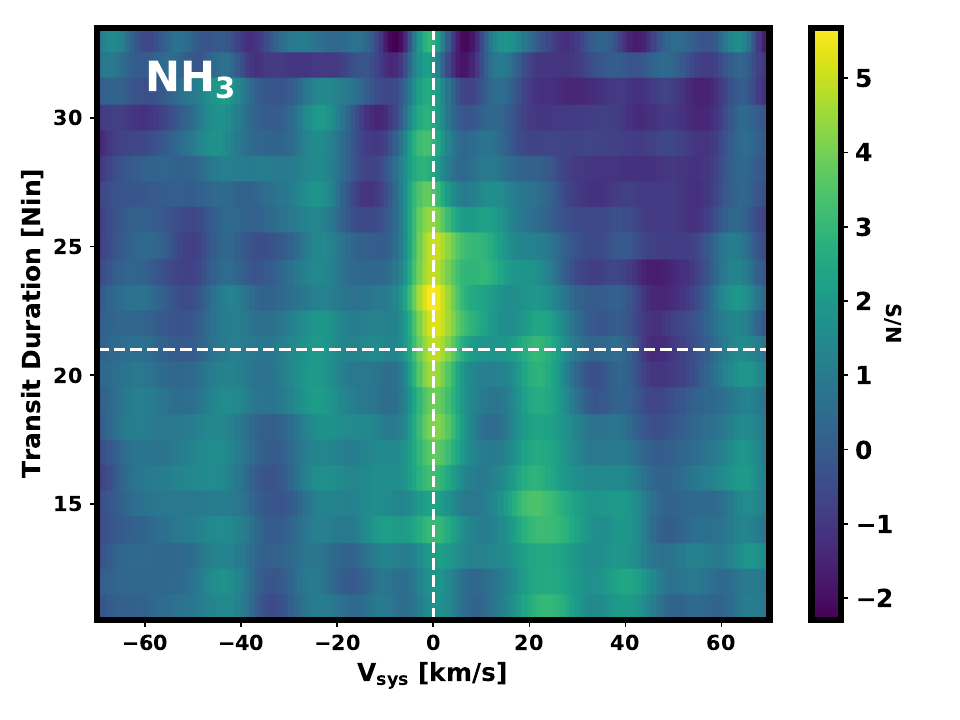}
    \includegraphics[width=0.5\textwidth]{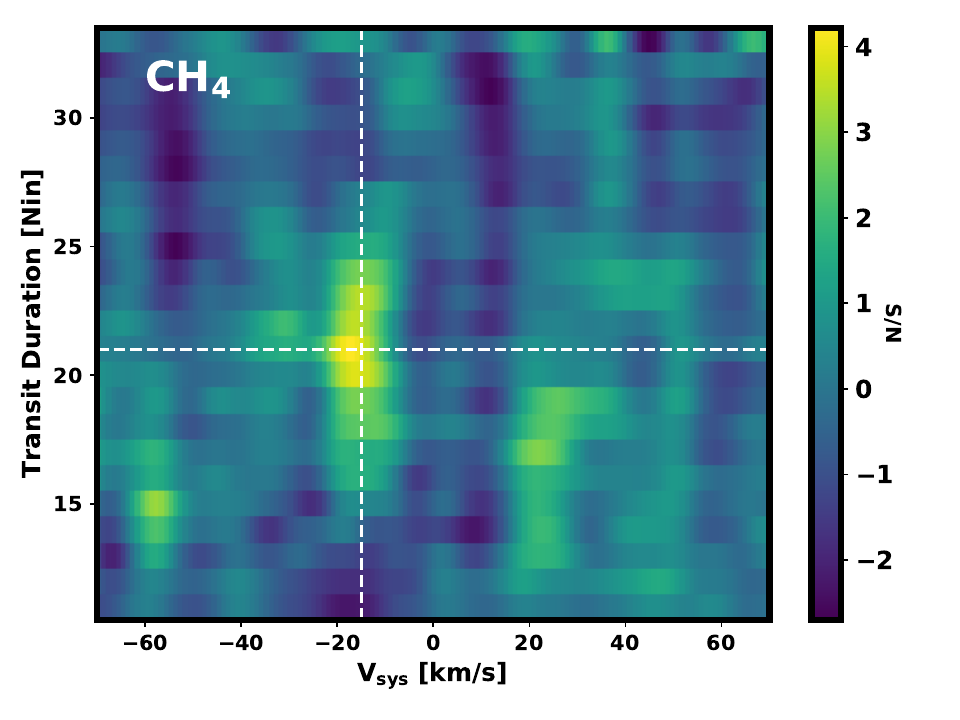}
    \includegraphics[width=0.5\textwidth]{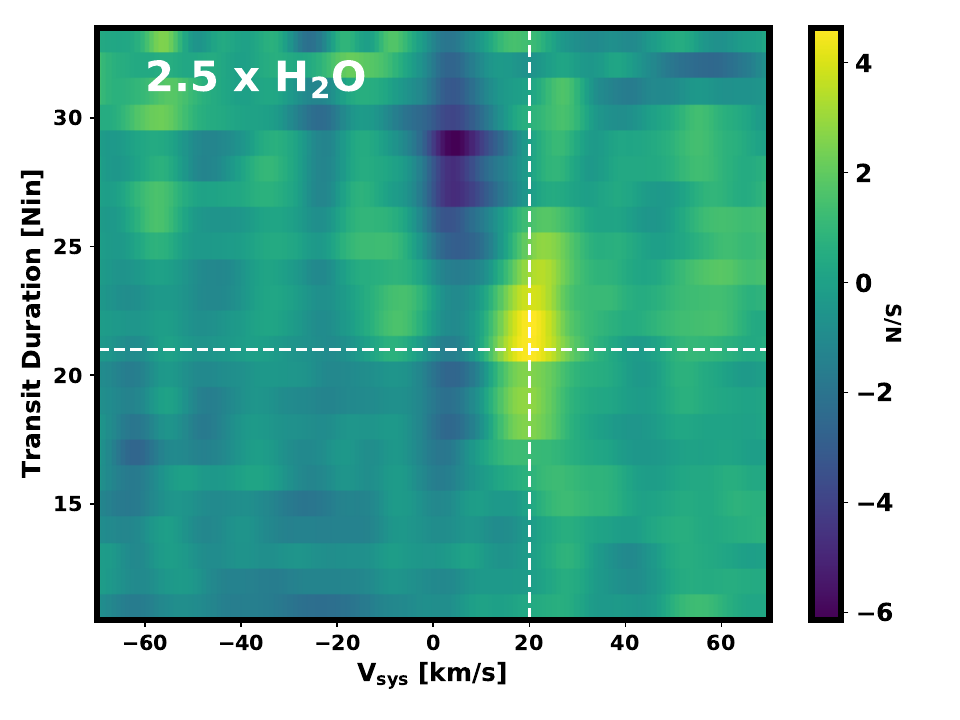}
    \caption{Constraining the injected signals in \nin - \vsys space. The resulting S/N across this space is shown when each of an \ce{NH3} (top), \ce{CH4} (middle), and an amplified \ce{H2O} (bottom) signal is injected into 21 in-transit spectra and then recovered, as in Figure \ref{fig:gamma_sn1}. For each point in this space the signals have been evaluated at the injection \kp~of 70 \kmsa. The signals are correctly constrained around \nin of 21, suggesting that they are indeed planetary rather than derived from telluric, stellar, or spurious features.}
    \label{fig:gamma_sn2}
\end{figure}
In Figure \ref{fig:gamma_sn2}, the cross-correlation S/N across \nin - \vsys space is shown for each molecule considered. Unlike for \kp, the recovered signals for each of \ce{NH3}, \ce{CH4}, and \ce{H2O} are constrained at the \nin, or \phimax, of injection. This suggests that the signals arise only during the transit, and are therefore likely to be planetary in nature. As well as this metric helping to distinguish being real planetary signals and stellar/telluric cross-correlation features, it may also aid in providing an independent measurement of the transit duration of the planet, in cases where this may not be well known or constrained. For example, for TOI-732 c studied here, values for $T_{14}$ vary across the literature, and with significant errors \citep{nowak2020, cloutier2020}.

However, a potential caveat of this method relates to the Rossiter-McLaughlin (R-M) effect \citep{rossiter_detection_1924, mclaughlin_results_1924, queloz_detection_2000}. The modulation of stellar signals through the R-M effect may produce spurious peaks constrained in $T_{14}$ space, which will therefore mimic planetary signals under this metric. For molecules that exist in the M-dwarf photosphere as well as in the planet's atmosphere, such as \ce{H2O}, this metric could therefore be unreliable.

\subsection{Importance of out-of-transit spectra in detrending}
\begin{figure*}
    \centering
    \includegraphics[width=0.49\textwidth]{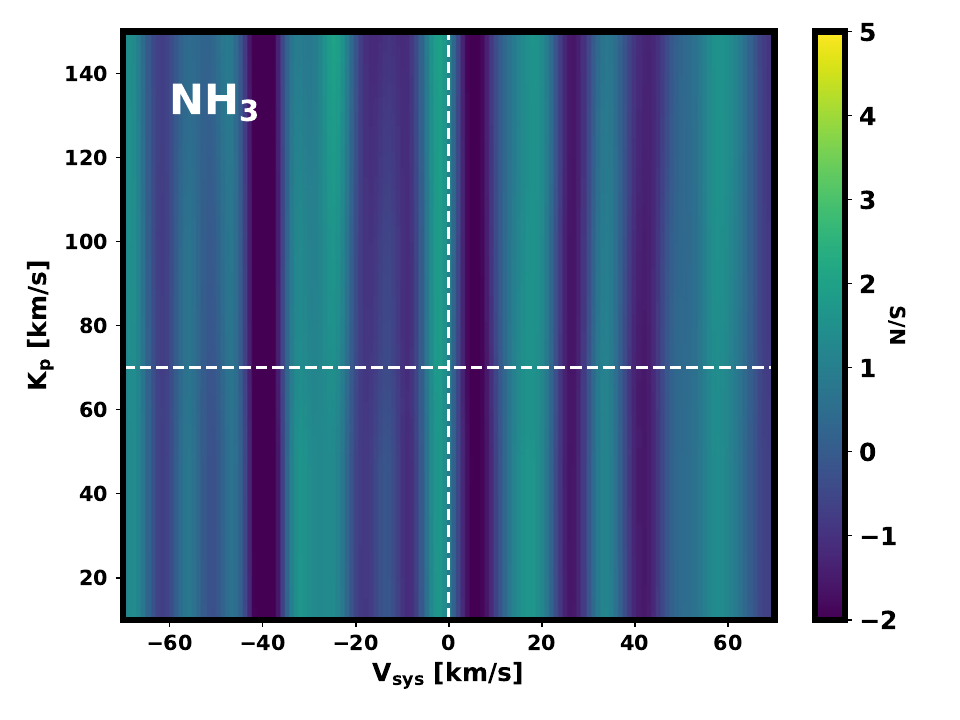}
    \includegraphics[width=0.49\textwidth]{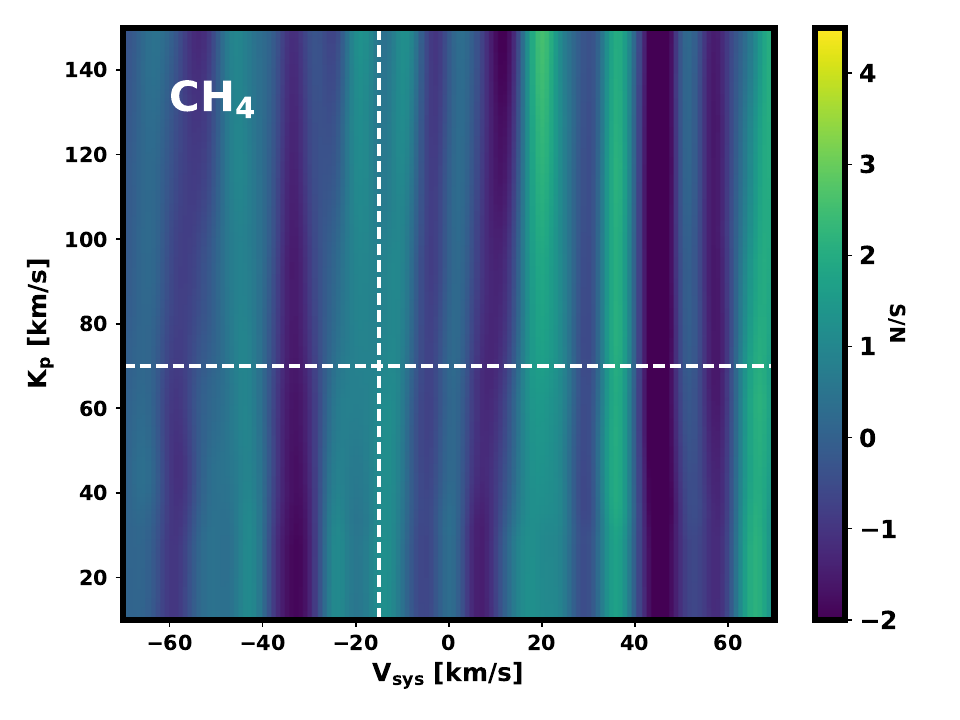}
    \caption{The low radial velocity change planet signals can no longer be recovered when there are no out-of-transit spectra used in the detrending. In each case the planetary signal is injected at \kp = 70\,\kmsa, as in Figure \ref{fig:gamma_sn2}, but the detection is no longer seen due to the loss in contrast between the planet signal and telluric/stellar contaminants. Left: for an injection of \ce{NH3} at \vsys of 0\,\kmsa, the 12 out-of-transit spectra are removed prior to detrending, leaving only the 21 in-transit spectra. Right: We inject a \ce{CH4} signal into all 33 spectra with a \vsys of -15\,\kmsa, such that the transit duration discriminator is removed. Again, the signal recovery is lost in this case.}
    \label{fig:results_no_out}
\end{figure*}
As discussed above, an important difference between the planet signal and the telluric/stellar residuals is that the planet signal is limited in time to just the in-transit spectra, whereas the others are present throughout our observations. In Figures \ref{fig:gamma_sn1} and \ref{fig:gamma_sn2}, the planet signals recovered were injected into 21 of 33 spectra (in accordance with the transit duration listed in Table \ref{toi732c_parameters}), with all 12 of the out-of-transit spectra used in the detrending. In the left hand panel of Figure \ref{fig:results_no_out}, we now trim our data set, in accordance with Section \ref{injection}, to remove all these out-of-transit spectra whilst the 21 in-transit spectra remain unchanged. In this manner, the injected \ce{NH3} planet signal is also unchanged, with the only difference being that now the out-of-transit spectra are not available for the fitting of principal components in detrending. With no out-of-transit spectra, we are unable to recover the \ce{NH3} planet signal, in direct contradiction to what is found in Figure \ref{fig:gamma_sn1}. Likewise in the right hand panel of Figure \ref{fig:results_no_out}, where the \ce{CH4} planet signal is injected into all 33 available spectra prior to detrending, the recovered signal is also lost. Given these results, it follows that it is very likely that it is indeed the out-of-transit spectra which allow us to successfully use PCA-based detrending for low-velocity targets; without these spectra the analysis is unable to recover the injected planetary signals. Although the planet signal is Doppler-shifted over the transit by an amount less than an instrument resolution element, the out-of-transit spectra provide contrast against the telluric and stellar lines which prevent the planet lines being included in the principal components of the time-series spectra. 
Without out-of-transit spectra, the planet signal is included in the principal components of the spectra and can therefore not be isolated. 
Considering frequencies, such contrast or time inhomogeneity may have the effect of increasing the signal frequency of the planet signal relative to the telluric/stellar contaminants. This could mean that sufficient planet signal is able to survive the subtraction of principal components, or perhaps the application of a time-series high-pass filter, during detrending.

We have therefore shown that it is indeed the contrast between the time-limited planetary signal and ubiquitous telluric and stellar features, provided by the out-of-transit spectra, that allows for the recovery of a low-velocity planet signal after detrending using PCA. In the high-velocity regime, the time-varying planet signal can still be recovered when the out-of-transit spectra are removed, albeit to a lower significance than in Figure \ref{fig:high_velocity} (e.g., S/N of 5.6 for \ce{NH3}).

\subsection{Effect of PCA on planet signal} \label{degredation}

\begin{figure}
    \centering
    \includegraphics[width=0.5\textwidth]{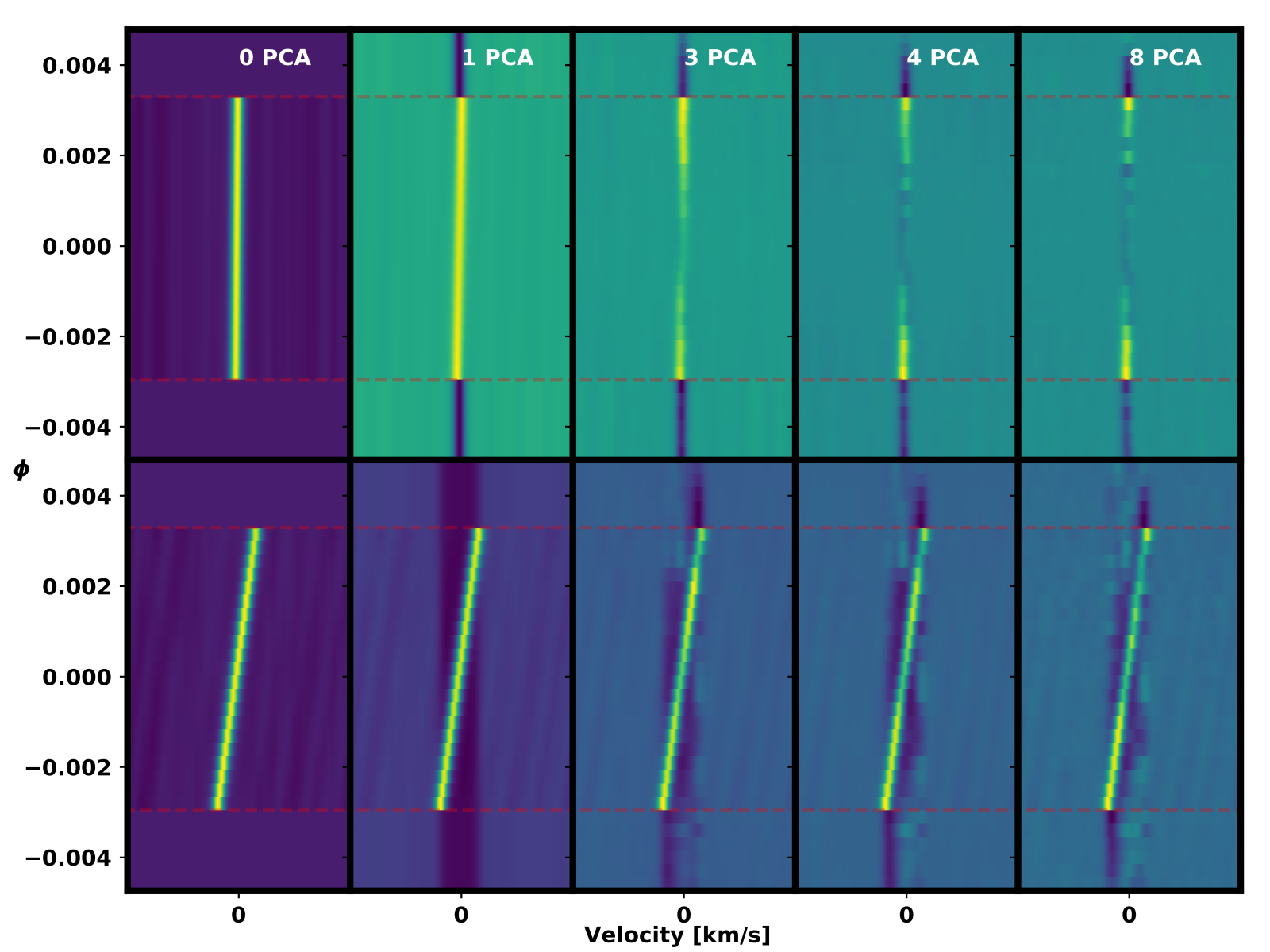} 
    \caption{The \deltaccf signal for \ce{NH3} after different numbers of principal components have been subtracted from the spectra. Top: For low-velocity planets, in this case \npx = 1.4, the main degradation of the planet signal occurs in the centre of the transit, with the edges preferentially surviving the successive subtraction of principal components. Strong out-of-transit artifacts are seen. Bottom: For high-velocity planets, e.g. a typical hot Jupiter with \npx = 25, the loss of planet signal after application of the PCA algorithm is slower and more equally distributed across the transit. Note that the top and bottom panels are scaled independently for clarity.}
    \label{fig:edge_concentration}
\end{figure}

In Figure \ref{fig:signal_degredation} it was shown that the planet signal degradation by the subtraction of principal components was faster for low-velocity planets than for warmer, short-period planets such as hot Jupiters.
However, we here note that the degradation of the planet signal is not consistent across the transit. This differential signal loss across the transit is demonstrated in Figure \ref{fig:edge_concentration}. We find that the ingress and egress points of the transit correspond to higher frequency components of the planet signal and are therefore degraded slower, and preserved for longer, after repeated subtraction of principal components from the spectra. On the other hand, the central part of the planet transit is included in the principal components of the spectra, and is therefore lost upon this subtraction. Therefore, after enough principal components are removed, the ingress and egress `edges' are the only surviving parts of the planet signal, and the available information about the planet is concentrated in these observations.
A similar effect can be seen when applying a simple high-pass filter to a upside-down, transit-like, top-hat signal. This results in the loss of signal apart from at the edges, where the extra variation corresponds to a higher frequency component of the signal.

Our finding suggests that in the low-velocity regime it is particularly important to reprocess the model to reflect this differential erosion by PCA, with the in-transit signal being removed by application of PCA significantly more than the ingress/egress signal. This is especially true when considering any time/phase-resolved spectroscopy of the planet, where constraints can be put on the variation of atmospheric conditions across different terminator viewing regions.

For the recovered signals shown in Figure \ref{fig:gamma_sn1}, only the in-transit spectra were included when summing the CCF over time. To further demonstrate the differential erosion of the planet signal in phase shown here, we now repeat the recovery of these injections using different portions of the data set. We find that including only the first and last 3 spectra of the transit, i.e. those spectra just after/before the ingress/egress, in the S/N calculation returns a value S/N $\gtrsim$ 3, whereas this threshold is not met when just the central 15 in-transit spectra are used, despite the increase in the number of in-transit spectra. We also calculate the S/N from the out-of-transit spectra only.
In this case, we recover a strong anti-correlation signal (S/N = -4.7) at \vsys of 0 \kmsa, indicative of the presence of the out-of-transit, in-trail artifacts observed in Figure \ref{fig:edge_concentration} and again suggesting that reprocessing is necessary (see Section \ref{reprocessing}). This unequal distribution of planet signal across the observations after detrending, characteristic of low-velocity planets, may be useful to consider when verifying real signals of such planets.

These effects seen here for low-radial-velocity-change planets may also mean that signals from such planets are potentially unsuitable for analysis with the Welch t-test \citep{welch_generalization_1947}, where a normal distribution in each of the in-trail samples is assumed. This metric would also likely be biased by the out-of-transit anti-correlated artifacts.

\section{Bayesian approach in the low-velocity regime} \label{likelihood_section}

The characterization of small and temperate planets, where any planetary signal is weak and slow-moving, could lend itself to more sensitive signal extraction methods. In this section we now repeat our analysis using a Bayesian log-likelihood framework. This may have a number of advantages over the cross-correlation metric. It has been shown in previous studies that a log-likelihood metric is ideally suited for model reprocessing \citep{brogi_retrieving_2019}, which our results suggest to be of particular importance for the characterization of low-velocity planets (Section \ref{degredation}). This may therefore mean that the likelihood framework is more sensitive to weak planet signals. Other advantages of a likelihood framework may include sensitivity to model strength, the ability to marginalise over unconstrained parameters, e.g. \kp, and avoidance of some of the difficulties associated with determining a detection S/N via cross-correlation, such as determining a velocity interval for noise estimation \citep{spring_black_2022, cheverall_robustness}. We here examine if our cross-correlation results for the characterization of low-velocity planets, presented in Section \ref{low_velocity_regime}, can be reproduced using this metric.

\subsection{Signal extraction using log-likelihoods} \label{signal_extraction_logl}

An alternative metric to cross-correlation by which the detection significance can be quantified is via a log-likelihood method \citep{brogi_retrieving_2019}. To implement this, the spectra are cleaned, normalized, and detrended as before. Rather than be cross-correlated with a Doppler-shifted model spectrum, the residuals are then compared with a reprocessed model template using a log-likelihood framework. In this way, a grid of atmospheric models of different chemical abundances, temperature-pressure profiles, Doppler-shifts, and scale factors can be compared against the data for goodness-of-fit. The formulation of the log-likelihood \logl for each trial model is described in the remainder of this section. 

\subsubsection{Model reprocessing} \label{reprocessing}

For an atmospheric model of a given chemical species, we shift the model template by a planetary velocity, given by Equation \ref{v_p}, for each point in a space spanned by \kp~and \vsys. The model is then multiplied by an atmospheric scaling factor $\alpha$. For the cross-correlation metric, where only line position and relative depth matter, this factor is set to one since the final detection S/N is agnostic to its value. However, its inclusion as a parameter in the log-likelihood metric makes this framework sensitive to information about line depths and shapes as well as positions \citep{brogi_retrieving_2019}. The extra constraint upon the recovered signal can allow for a more confident detection, acting as an identifier which can further distinguish spurious signals from real planet signals ($\alpha \approx 1$ for the correct model).

Before comparison to the residual data, we now reprocess our models to reproduce the effects of the PCA-based detrending on the real planet signal \citep{brogi_retrieving_2019}. This is an important step \citep{brogi_roasting_2022, lafarga2023}, with the planet signal being significantly `eroded' and distorted with each principal component removed (see Figure \ref{fig:signal_degredation}), which can potentially lead to detection biases \citep{brogi_retrieving_2019}. To do this model reprocessing, we follow the approach in \cite{brogi2023_marsh}.
We first take the sum of the removed principal components calculated from the data, i.e. what was subtracted in detrending, and inject into this empirically-derived telluric template the Doppler-shifted and scaled planet model. We then run the PCA algorithm again in the same way as before. 
This is then subjected to the same PCA detrending as the original spectra. The reprocessing, done as it is here by injecting a model planet signal into the removed principal components, which represent the telluric and stellar spectra in addition to other correlated noise sources, e.g. from the instrument, is also able to reproduce the modulation that these contaminating signals may have on the planetary transmission spectrum. Note that we do not here model or test the mitigation that this metric may provide against spurious signals generated by the R-M effect, as discussed in Section \ref{novel}.

The reprocessing of the model to replicate the effects of the PCA-based detrending on the real planetary signal is likely to give a closer match between the model and the remaining signal after several principal components have been removed, thereby leading to an increase in detection sensitivity. In Figure \ref{fig:edge_concentration}, it can be seen that the application of PCA on a planetary transit signal produces residual anti-correlation features in the out-of-transit regime. If cross-correlating the in-transit spectra with a planet model, as was done in Sections \ref{methods} and \ref{low_velocity_regime}, then this residual signal will not be recovered and will contribute to the losses observed in Figure \ref{fig:signal_degredation}. However, a model comparison framework which suitably reprocesses the planet model template may be able to capture these residuals in the signal summation, and hence return a greater detection significance.

\subsubsection{Model comparison} \label{model_comparison}

The resulting planet atmosphere model, reprocessed to account for the effects of detrending, can then be directly compared with the residual data to calculate a log-likelihood value at one point in \kp - \vsys - $\alpha$ space. Whilst a cross-correlation to log-likelihood mapping can be used \citep{zucker2003, webb_water_2022, lafarga2023}, we here calculate \logl directly using the following equation (also see \cite{brogi_retrieving_2019, gibson2020}):
\begin{equation} \label{logl_calc}
\mathrm{log}(L)(K_{\mathrm{p}},V_{\mathrm{sys}},\alpha) = -\frac{1}{2} \sum_{i, j}\left[\left(\frac{r_{ij} - m_{ij}}{\sigma_{j}}\right)^2 + \mathrm{log}(2\pi\sigma_{j}^2)\right] 
\end{equation}
where $r_{ij}$ is the residual spectrum after detrending at wavelength channel $j$ and observation $i$, $m_{ij}$ is Doppler-shifted and reprocessed model spectrum at the corresponding wavelength and observation, and $\sigma_{j}$ is the error in the residuals for wavelength channel $j$. %
We have here assumed that for any given pixel the variance is contributed by the temporal variation of the flux throughout the observations. This is estimated for each wavelength channel using the median absolute deviation of the detrended residuals.

We repeat this comparison across model parameter space, with the resulting log-likelihood distribution used to calculate a Bayesian detection significance, as described in Section \ref{bayes_detection_significance}.

\subsection{Bayesian detection significance} \label{bayes_detection_significance}

In addition to the log-likelihood distribution across parameter space for a model of a given chemical species, the log-likelihood value is calculated for a null model not containing that species. We here use a flat, zero transit-depth spectrum (no planet transit: $\alpha$ = 0) as the null model. A Bayesian evidence, or marginal likelihood, is then calculated for each of the atmospheric and null models, from which a detection probability/significance can be determined. To calculate this Bayesian evidence, the log-likelihood distribution of the atmospheric model is integrated over the model parameters, modulated by their prior probability distributions:
\begin{equation} \label{evidence_calc}
Z_{m} = \int_{\theta} log(L)\pi(\theta) d\theta
\end{equation}
To obtain a Bayesian detection significance, the overall model evidence is first calculated as above.
A Bayes factor $B$ is then calculated as the ratio of the atmospheric model evidence to the null model evidence, $Z_{\mathrm{m}}/Z_{0}$. This factor is converted to the detection significance $\sigma$ of the atmospheric model relative to the null model using the following expression, from \cite{welbanks2021} based on work by \cite{benneke2013}:
\begin{equation} \label{DS_calc}
\sigma = \sqrt{2}.\mathrm{erfc}^{-1} \left[\mathrm{exp} \left(W_{-1} \left(- \frac{1}{Be} \right) \right) \right]
\end{equation}
where erfc is the complementary error function and $W_{-1}$ is the Lambert $W$ function in its lower branch ($k = -1$).

The ability to integrate over parameters to form a marginal likelihood is useful in the low-velocity planet regime, where \kp~is found to be unconstrained (Section \ref{uncertainty}). This may enable more consideration of the signal's degeneracy compared to cross-correlation, where in Section \ref{novel} we fixed \kp~ at its expected value when exploring \nin space.

\subsection{Application to \ce{H2O} signal injection}

We now apply this metric to our data set, as we did previously using cross-correlation, to recover an injected \ce{H2O} signal. We use this model planet signal as a conservative test case since it was recovered weakly with S/N = 2.1 using the cross-correlation method in Figure \ref{fig:gamma_sn1}. We again detrend the spectra by subtracting 7 principal components. To compare the spectral residuals to the model, we calculate log-likelihood values across the 3-dimensional model space spanned by \kp, \vsys, and $\alpha$.

\begin{figure}
    \centering
    \includegraphics[width=0.49\textwidth]{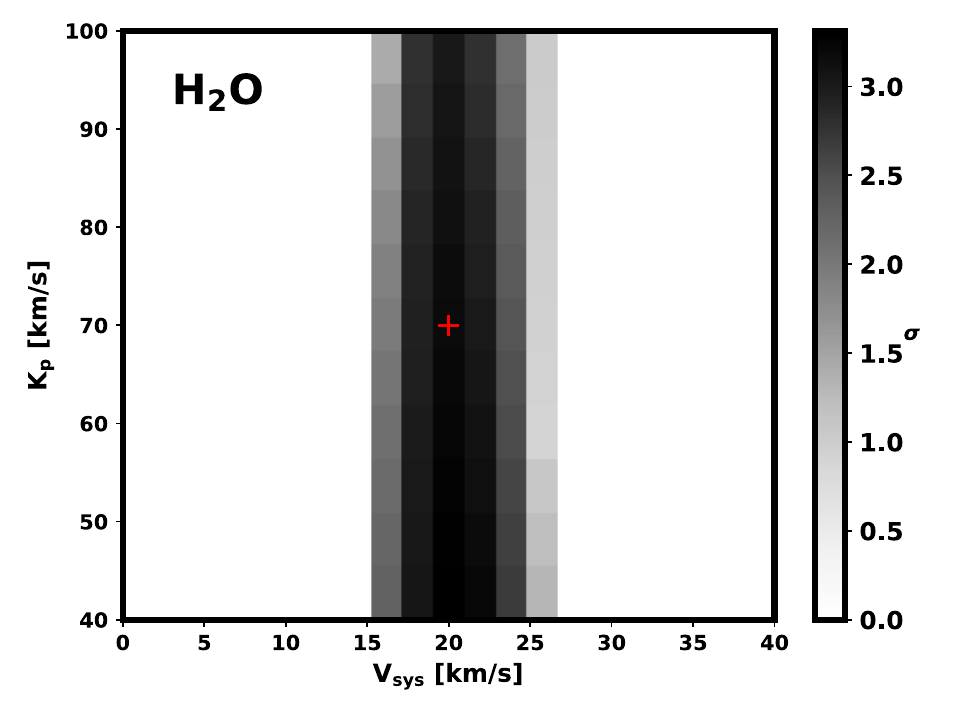}
    \includegraphics[width=0.49\textwidth]{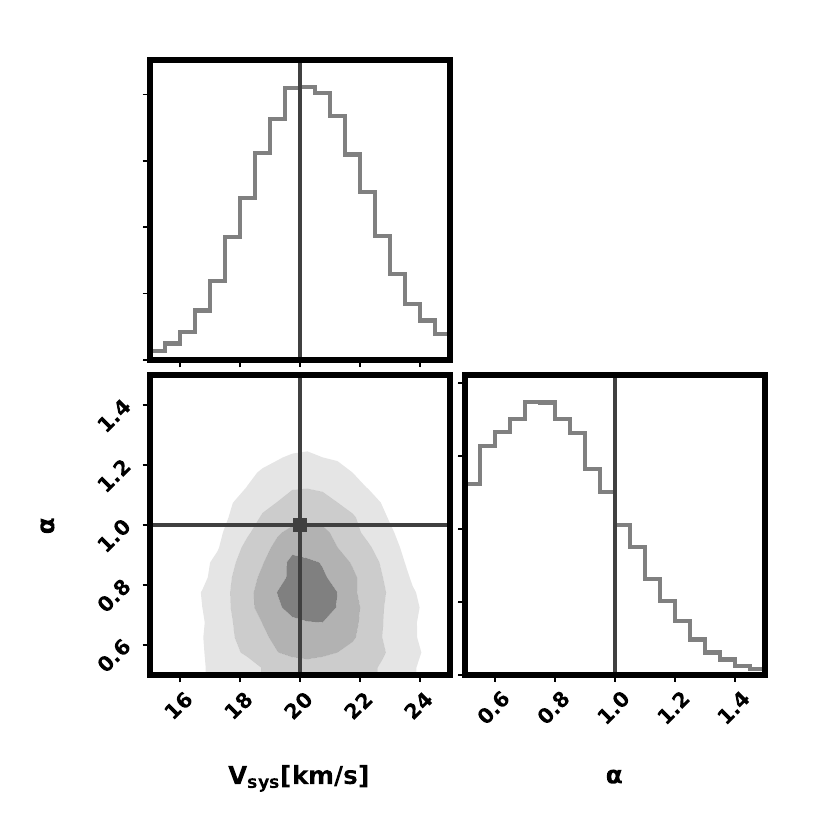}
    \caption{The low-velocity \ce{H2O} planet signal, initially seen using cross-correlation (see the right-hand panel of Figure \ref{fig:gamma_sn1}), is now recovered using a Bayesian log-likelihood approach. As before, the planet model was injected into the data with \kp = 70\,\kms and \vsys = 20\,\kmsa, for 21 in-transit and 12 out-of-transit spectra. Top: By comparing the detrended spectral residuals with a reprocessed model, and marginalising over $\alpha$, a detection significance is calculated across \kp - \vsys space. Bottom: the posterior probability distribution across \vsys-~$\alpha$ space is shown, with values of \vsys = $20\pm2$ \kms and $\alpha = 0.8\pm0.2$ found.  In both cases, a signal is recovered with parameters consistent with the injection (marked with a red cross/crosshairs). A final detection significance of 2.9$\sigma$ is achieved for \ce{H2O} compared with the null model. Note that this is smaller than the peak significance shown in the top panel, where only $\alpha$ has been marginalised, since we are no longer assuming perfectly known \kp and \vsys at each point.}
    \label{fig:logl}
\end{figure}

The recovered signal is shown in Figure \ref{fig:logl}. In the top panel, the likelihood function is marginalised over just $\alpha$ (using uniform priors of 0.5 to 1.5) to give the model evidence and then Bayesian detection significance as a function of \kp~and \vsys. In the bottom panel, we show the posterior probability distribution across \vsys-~$\alpha$ space calculated from our likelihood distribution using \textit{emcee} \citep{emcee2013}. In doing so we use the above prior for $\alpha$ and uniform priors of 40 to 100 \kms for \kp and 15 to 25 \kms for \vsys. Values of \vsys = $20\pm2$ \kms and $\alpha = 0.8\pm0.2$ are found.
Clear preference is shown for an atmospheric model with parameters consistent with the injection. Marginalising all the model parameters over the entire uniform prior volume, as described in Section \ref{bayes_detection_significance}, results in a detection significance of 2.9$\sigma$ for \ce{H2O} compared with the null model.
As in the cross-correlation case, no significant signals are recovered without an injection, demonstrating the successful PCA-based extraction of the injected planet signal from spectra of a low-velocity, temperate planet.
Our results may therefore suggest that the Bayesian framework used here is more sensitive to weak planet signals than the cross-correlation metric used in Section \ref{low_velocity_regime}, for example due to model reprocessing and the extra sensitivity to line depths and shapes.
Once again, removing all 12 out-of-transit spectra prior to detrending means the injection is no longer recovered.

\section{Observational limits and trends in low-velocity regime} \label{limits}

In Sections \ref{low_velocity_regime} and \ref{likelihood_section} we have shown that it is possible to probe the atmospheres of low-velocity planets, such as TOI-732 c, using PCA-based detrending. In this section we use the S/N metric to demonstrate trends and investigate further the limits of high-resolution transmission spectroscopy as applied to low-velocity planets. We use the same spectra and methods as before to examine the lowest velocity orbits for which it is possible to achieve a detection in high-resolution transmission spectroscopy, and illustrate once more that under the correct observing conditions it is possible to recover planetary signals down to sub-instrumental resolution velocities. For low-velocity planets, we find that there are two main factors which primarily determine our ability to sufficiently isolate a given planetary signal from the observed spectra. We outline these throughout this section. 
Throughout this section we inject our model signals before normalisation, for ease of sampling a large injection velocity parameter space, and the similarity in results found previously.

\subsection{Planetary signal pixel shift} \label{shift_section}

PCA-based detrending uses changes/inhomogeneities in the time-series spectra, such as the change in the planetary radial velocity during the transit or the existence of transit edges, to differentiate the Doppler-shifting/temporary planetary lines from the quasi-static and persistent telluric and stellar lines. We here investigate how the quality of detrending, reflected by the recovered S/N of an injected signal, is dependent on the number of instrument wavelength channels by which the planetary signal is Doppler-shifted during the transit, \npxa, as introduced in Section 3.5. This is dependent on \kp, the maximum phase during transit \phimax, and the instrument pixel velocity \vpx, as defined in equation \ref{npx_equation}.

\begin{figure}
    \centering
    \includegraphics[width=0.5\textwidth]{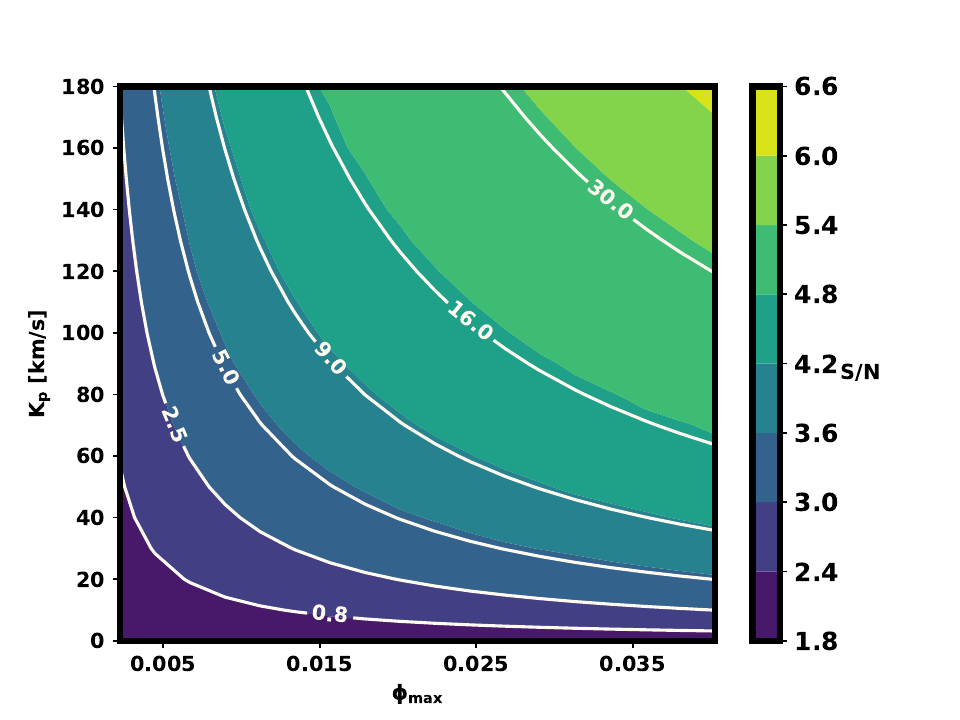}
    \caption{The extent to which a planet signal can be recovered increases with the planet pixel shift, \npx, or number of pixels across which the planetary signal is spread. The detection S/N of a \ce{CH4} signal, injected with a \vsys of -15 \kms into 16 in-transit spectra, with 12 out-of-transit spectra, and detrended by subtracting 3 principal components, is shown as a function of \kp - \phimax~space. Contours showing lines of constant \npx, calculated using the resolution of the IGRINS data used here, are shown, aligning strongly with the increase in S/N. This trend begins to slow at higher values of \npx. Signals with changes in radial velocity less than that corresponding to one instrument resolution element can still be recovered. The barycentric velocity correction is here set to zero, and due to the even number of in-transit spectra simulated, \phimax~here represents half the in-transit phase range of each simulation.}
    \label{fig:npx}
\end{figure}

We inject and recover \ce{CH4} models over a range of simulated \kp~and \phimax, for constant \vsys and numbers of in-transit and out-of-transit spectra (-15\kmsa, 16, and 12, respectively). The S/N of the recovered signal is plotted across this \kp - \phimax~space in Figure \ref{fig:npx}, with \npx contours, calculated from these orbital parameters and the resolution of the IGRINS data used here, shown. This figure illustrates the strong dependence of the detrending on the planetary pixel shift, and shows how it can be useful to consider the compound parameter of \npx alone, rather than each of \kp, \phimax, and \vpx individually, when predicting whether a given detection can be made. As expected, in the low orbital velocity regime, the recovery of a signal increases with increasing \npx. As we approach greater values of \npx (warmer planets with larger changes in radial velocity during transit), the increase in detection S/N begins to slow, after which more time-variation in the planet signal leads to diminishing returns. We once again find that planetary signals can still be recovered when the change in the planet's radial velocity is small, including when it is smaller than that corresponding to an instrument resolution element, i.e. \npx $\lesssim$ 3.

\subsection{In-transit / out-of-transit observation ratio} \label{ratio_section}

We have shown that it is possible to probe the atmospheres of slower moving planets using PCA-based detrending due to the contrast provided by the out-of-transit spectra.
We now investigate further how the number of out-of-transit spectra affects our ability to isolate a planet signal from the telluric/stellar contaminants using PCA-based detrending. As before, we do this by measuring the S/N at which various injected signals are recovered.

\begin{figure*}
    \centering
    \includegraphics[width=0.49\textwidth]{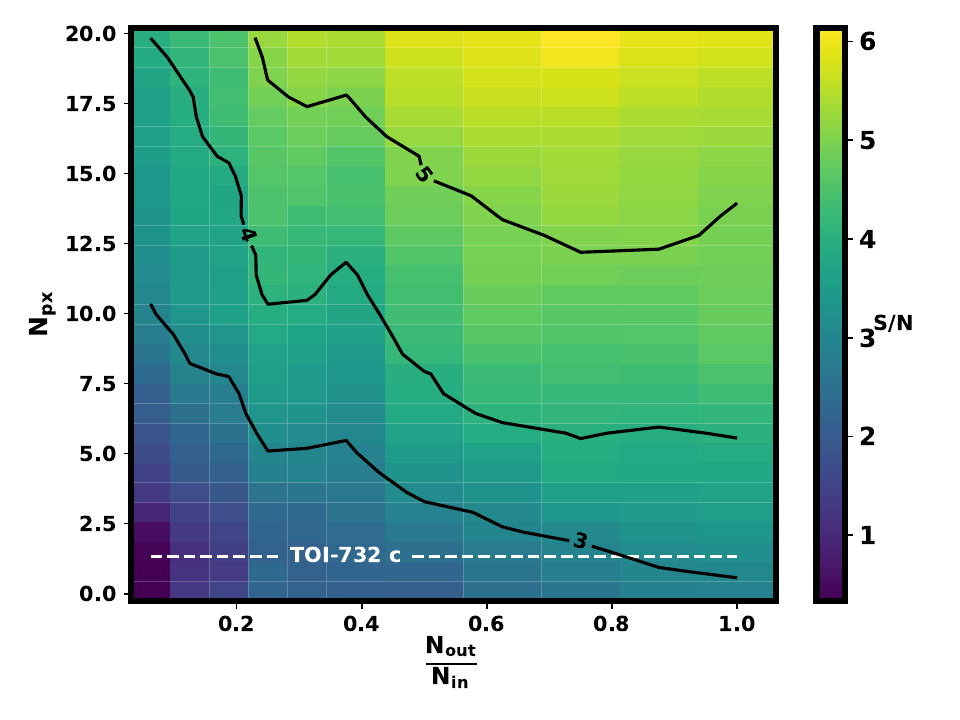} 
    \includegraphics[width=0.49\textwidth]{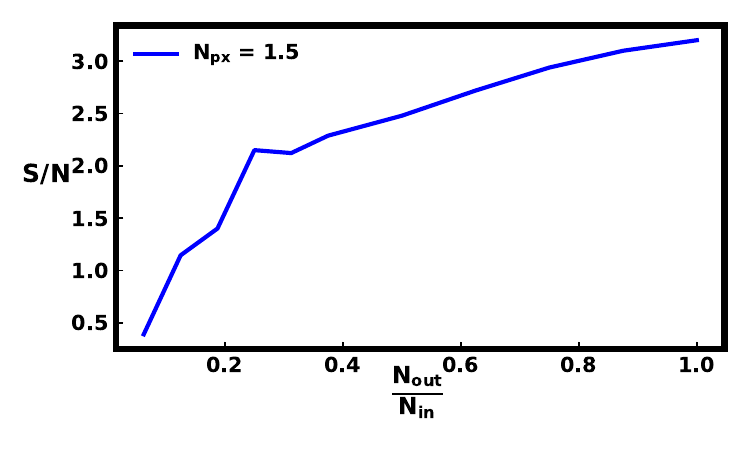} 
    \caption{We find that the extent to which planet signals can be isolated from the observed spectra is driven by two factors: the planet pixel shift, \npx, and the ratio of out-of-transit spectra to in-transit spectra, \nout/\nin. Left: The S/N across the space spanned by these two parameters is shown, for the case of recovering a \ce{CH4} model injected with \nin = 16, using 5 PCA iterations. The S/N given is the median found when sampling injections over a large number of possible \vsys. Increasing ability to detrend, represented by an increase in recovered S/N, is demonstrated when increasing these parameters. With few out-of-transit spectra, it is very difficult to recover low-velocity (\npx $\lesssim$ 3) signals. The \npx value for the planet and instrument in this study, TOI-732 c, without the barycentric contribution is marked in white (\npx = 1.3; Table \ref{rv_change}). In the case shown here, a significant signal can be recovered for this planet and model when the number of out-of-transit spectra $\gtrsim$ 0.8 x the number of in-transit spectra. Right: For the case of \npx = 1.5, similar to that of TOI-732 c, the median S/N across \vsys is shown against the out-of-transit to in-transit ratio. A significant increase in S/N is found as more out-of-transit spectra are observed.}
    \label{fig:K_Nout}
\end{figure*}

For a fixed number of in-transit spectra, we vary the number of out-of-transit spectra included in the detrending. Given that in Section \ref{shift_section} we demonstrate the importance of the planetary pixel shift, Figure \ref{fig:K_Nout} shows the regions of the space spanned by \npx and the ratio of out-of-transit spectra to in-transit spectra which are suitable for obtaining a detection in high-resolution transmission spectroscopy. A \ce{CH4} model injected into 16 in-transit spectra is used here, and the S/N values presented in this figure are evaluated by averaging over \vsys space. We demonstrate that the recovery of a given signal increases when a greater fraction of out-of-transit spectra are included in the detrending. This is expected as the existence of the planet signal in a smaller fraction of the data creates more contrast with the tellurics and so means less of the signal is included in the principal components of the time-series spectra. The effect is to promote a more effective separation of telluric and planetary signals; even for very small values of \npx, atmospheric characterisation can be achieved if sufficient out-of-transit spectra are included in detrending. The planetary pixel shift of TOI-732 c is marked, calculated using the values for \kp~and \phimax in Table \ref{toi732c_parameters}, and assuming an instrument resolution equal to that of IGRINS (\vpx $\approx$ 2 \kmsa). In the case of these simulations we find that, averaged over \vsys, a TOI-732 c -like sub-Neptune planet signal can be recovered with S/N > 3 when the ratio of out-of-transit spectra to in-transit spectra is $\gtrsim$ 0.8. This trend could therefore be an important consideration when planning observations of low-velocity planets. 

\begin{figure}
    \centering
    \includegraphics[width=0.5\textwidth]{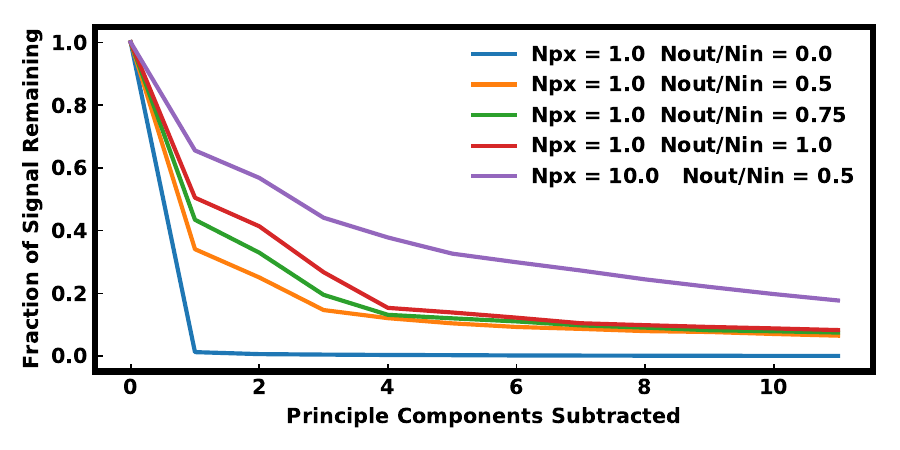}
    \caption{The degradation of the planet signal by PCA varies for different numbers of out-of-transit observations. Similar to Figure \ref{fig:signal_degredation}, an \ce{NH3} signal is used here, with the rate of signal removed with each principal component subtracted explored across $N_{\mathrm{out}}$/$N_{\mathrm{in}}$ space. With no out-of-transit spectra, the signal is effectively entirely removed after 1 iteration of PCA, consistent with Figure \ref{fig:results_no_out}. Observing more out-of-transit spectra slows this removal, giving the trend in S/N seen in Figure \ref{fig:npx}. A higher velocity signal is also shown for comparison.}
    \label{fig:signal_degredation_nin_nout_npx}
\end{figure}
We now explore the erosion of a low-velocity \ce{NH3} planet signal with each principal component subtracted from the spectra, as in Figure \ref{fig:signal_degredation}, for different out-of-transit to in-transit ratios.
The number of in-transit spectra is again set as 16 in this test, with a higher velocity signal also shown for comparison. As expected from Figure \ref{fig:K_Nout}, the planet signal is lost more quickly when there are fewer out-of-transit spectra.
In the case of no out-of-transit spectra and a sub-resolution change in radial velocity, the remaining planet signal is negligible after just one iteration. When more out-of-transit spectra are included in detrending, or even when there are no out-of-transit spectra but the change in the planet's radial velocity is larger, more of the planet signal evades inclusion in the principal components of the spectra. In these cases the signal is still able to be recovered after a number of principal components have been subtracted to remove sufficient correlated noise.

\section{Partial transits and extension to emission spectroscopy} \label{partial_emission}

We have established that for a given planet, the presence of out-of-transit spectra can help preserve the in-transit planet signal upon the subtraction of principal components during detrending. We now investigate how the set-up of our observations, e.g. the symmetry of out-of-transit spectra, can affect our findings.

\subsection{Partial transits} \label{partial}

Sometimes only partial transits are available; conditions e.g. airmass may not permit the observation of a full transit plus both pre-transit and post-transit spectra. In this case the number of pre-transit and post-transit observations will likely be unequal. There may be out-of-transit spectra available on only one side of the transit, or perhaps only a fraction of the transit itself was observed. In this section we explore the case where there is an unequal number of pre-transit and post-transit observations. Until now we have injected signals into in-transit spectra and selected the out-of-transit spectra symmetrically in phase, as detailed in Section \ref{injection}. Since our phases are symmetrical about zero, this approach has led to roughly equal numbers of out-of-transit spectra on each side of the transit.

\begin{figure}
    \centering
    \includegraphics[width=0.5\textwidth]{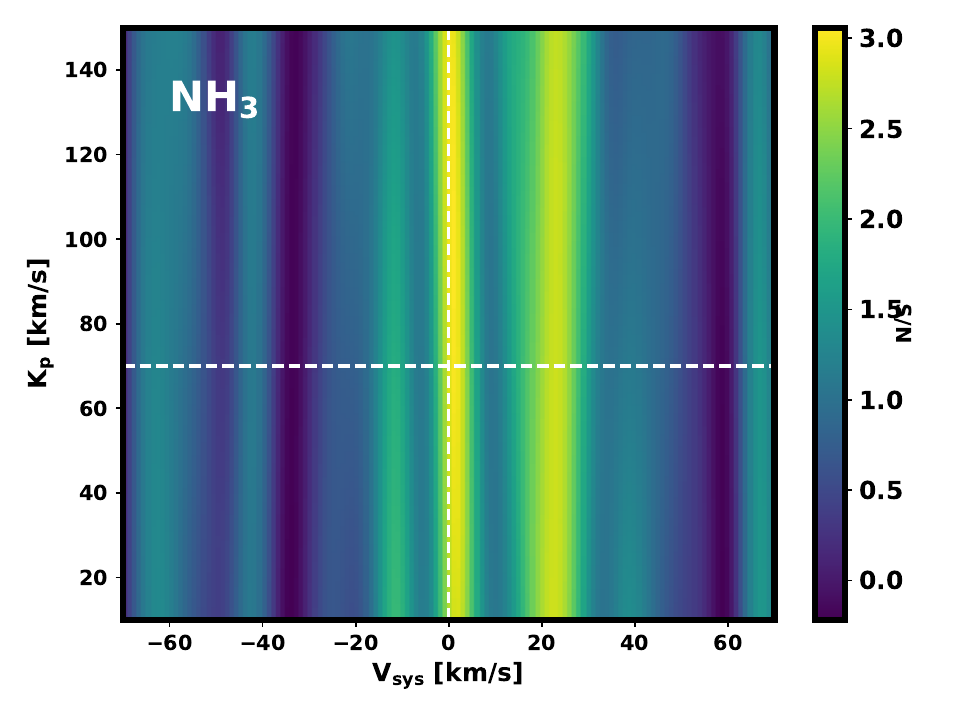}
    \includegraphics[width=0.5\textwidth]{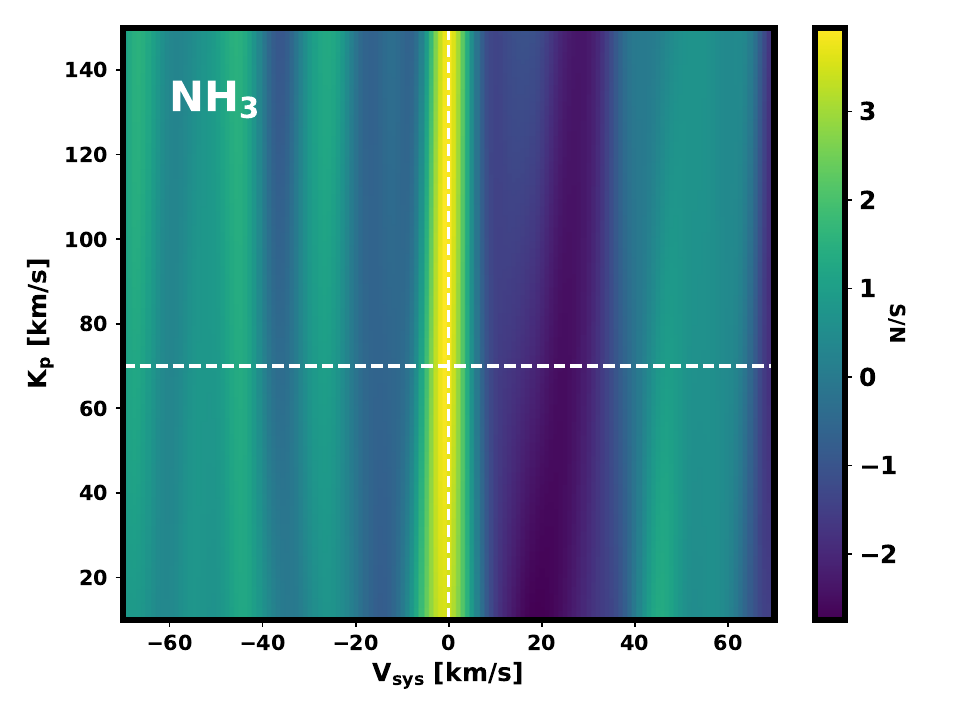}%
    \caption{Low-velocity planet signals can still be recovered from partial transits although to a lower significance. In each case shown here, the signal has been injected into 21 in-transit spectra with a planetary velocity of \kp = 70\,\kmsa. There are 12 out-of-transit spectra in both plots, but whereas in Figure \ref{fig:gamma_sn1} these are split equally between pre- and post-transit, here they are all either before or after the transit. Top: for 12 pre-transit spectra, the signal is recovered with S/N of 3.1. Bottom: for 12 post-transit spectra, the signal is recovered with S/N of 3.9.}
    \label{fig:unequal1}
\end{figure}

We complete injection and recovery tests as in Figure \ref{fig:gamma_sn1}, this time simulating the case where all out-of-transit spectra are either before or after the 21 in-transit observations (Figure \ref{fig:unequal1}). We find that whilst we are still able to recover signals from partial transits, it can be more difficult, demonstrated by the reduced S/N values found here for NH3 (3.1 and 3.9) compared to the value from Figure \ref{fig:gamma_sn1} (5.0). Note the greater S/N in the second partial transit example (bottom panel, with all the out-of-transit observations after the transit), likely due to the higher airmass towards the ends of our observations.
In the case of a partial transit, the lack of out-of-transit spectra on one side of the spectrum removes an `edge' from the transit profile, further reducing the frequency of the planet signal and making it more difficult to separate from the telluric, stellar, and other sources of correlated noise (see Section \ref{degredation}). 
It therefore is preferable to have both sufficient pre- and post-transit observations.

\subsection{Emission spectroscopy} \label{emission}

We note that our work here has focused solely on transmission spectroscopy. In emission spectroscopy, the typical existence of the planet signal in every observation means it is removed by PCA during detrending if the radial velocity change is too small. Despite this, our findings here suggest that it may be technically possible to detrend emission spectra in this regime if observations of the system include some of the secondary eclipse. These in-eclipse spectra, taken before/after/between observations in which the planet is visible, could then perhaps act as the contrasting `out-of-transit' spectra required for the detrending to work. In this case, the set up reduces to case of partial transits, with `out-of-transit' observations taken only before/after the transit. Whilst we note that temperate planets will not typically have detectable emission features, and therefore their lack of radial velocity change during observations is unlikely to be the main obstacle, this idea may find relevance in the study of young giant planets.

\section{Summary and discussion} \label{summary}

In this work we investigate the feasibility of atmospheric characterisation of long-period low-mass exoplanets orbiting M dwarfs using high-resolution transmission spectroscopy. We show that it is indeed possible to make chemical detections in the atmospheres of such planets using high-resolution transmission spectroscopy and PCA-based detrending, even at sub-pixel changes in radial velocity and small systematic velocities. This is because, despite the small change in planetary radial velocity during the transit failing to differentiate the planet signal from the telluric and stellar lines, the contrast provided by the out-of-transit spectra prevents the planet signal from being included in the principal components subtracted in detrending.

We find that there are two main observational drivers to the quality of PCA-based detrending: the number of instrument wavelength channels by which the planetary signal is Doppler-shifted during the transit (referred to here as the planet pixel shift), and the ratio of the number of out-of-transit spectra to in-transit spectra. We have shown that both of these factors increase the detection S/N. Increasing the resolution of high-resolution spectrographs will result in the spectral lines of a given planet crossing more instrument wavelength channels during the transit, but instrument limitations prevent this from being the solution for the slowest moving planets. Instead, careful selection of the observational set-up, particularly ensuring that a sufficient out-of-transit to in-transit observation ratio can be achieved, can aid in detrending. This will allow us to study the compositions of temperate planetary atmospheres. Other considerations, such as timing the observations to coincide with periods of maximally changing barycentric velocity correction and of low stellar activity \citep{sairam2022}, could also be relevant as we push the limits of long-period, low-mass exoplanets.

We note that the trends and limits found in this work make use of only one data set of a specific target, alongside a small selection of models. However, we expect that the general trends and key findings uncovered should hold true for the majority of data sets. Whilst our method of injecting a planetary signal into a real data set provides increased realism over wholly simulated spectra, it is also limited in some regards. For example, we are not able to vary the Doppler-shift of the stellar spectrum contained in the data when injecting planet signals at different systematic velocities.

Extending the regime of planets possible to characterize with ground-based high-resolution transmission spectroscopy to those with longer periods will allow us to probe habitable zone terrestrial and sub-Neptune/Hycean-candidate planets around M-dwarf stars \citep{madhusudhan2021}. Once sufficient S/N can be achieved, e.g. with future extremely large telescopes, to observe the thin, high mean molecular weight atmospheres that may exist on terrestrial planets, e.g. super-Earths, then ground-based observations may also be able to search for potential biosignatures using the approach presented here. We have extended the parameter space of planets accessible for chemical characterisation from the ground and paved the way towards the first high-resolution spectroscopic detection of a molecule in the atmosphere of a long-period, temperate sub-Neptune around an M-dwarf.

\begin{acknowledgments}
This work is supported by research grants to N.M. from the MERAC Foundation, Switzerland, and the UK Science and Technology Facilities Council (STFC) Center for Doctoral Training (CDT) in Data Intensive Science at the University of Cambridge. N.M. and C.C. acknowledge support from these sources towards the doctoral studies of C.C. C.C. would also like to thank Måns Holmberg and Sam Cabot for useful discussions. We thank the team of Gemini GO program GS-2021A-Q-201 (PI: D. Valencia) and the IGRINS team at Gemini-S for conducting the observations and data reduction. NM thanks Gregory Mace of the IGRINS team for helpful discussions. 

This research has made use of the NASA Exoplanet Archive, which is operated by the California Institute of Technology, under contract with the National Aeronautics and Space Administration under the Exoplanet Exploration Program. This research has made use of the NASA Astrophysics Data System and the Python packages \texttt{NUMPY}, \texttt{SCIPY}, and \texttt{MATPLOTLIB}.

\textit{Author Contributions}: N.M. conceived and planned the project, contributing to the successful observing proposal, facilitating the observations, and conducting the atmospheric modelling. C.C. conducted the data analysis and led the writing of the manuscript with comments and edits from N.M.

\end{acknowledgments}

\bibliography{paper_file}{}
\bibliographystyle{aasjournal}

\appendix

\begin{figure*}[h]
    \centering
    \includegraphics[width=0.33\textwidth]{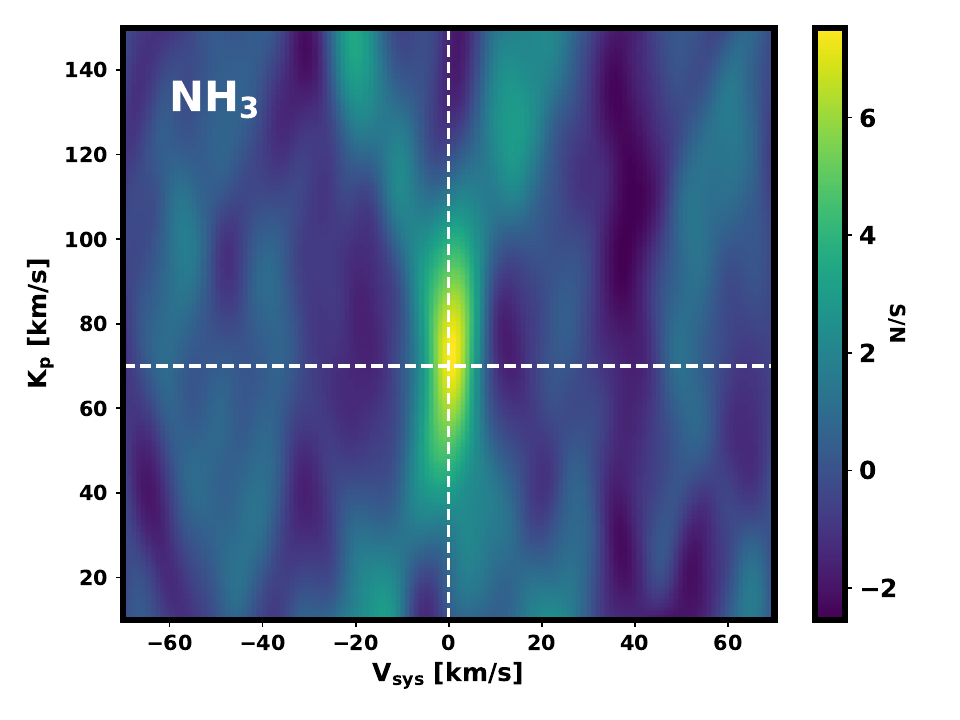}
    \includegraphics[width=0.33\textwidth]{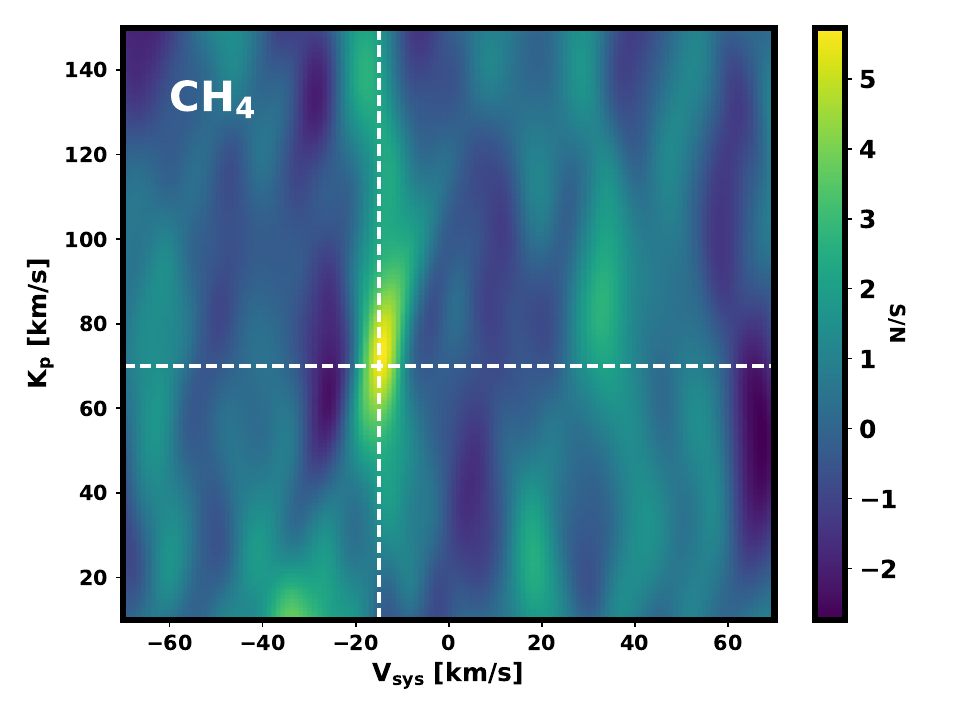}
    \includegraphics[width=0.33\textwidth]{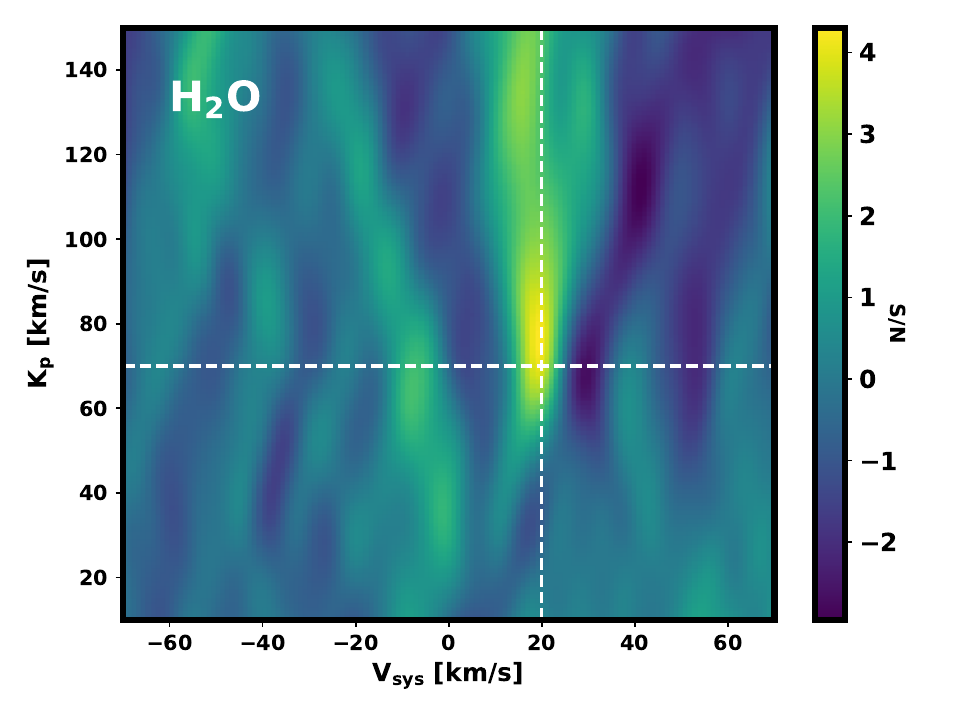}
    \caption{S/N maps for high-velocity injection and recovery tests, for comparison to those for the low-velocity injection tests shown in Figure \ref{fig:gamma_sn1}. The injected \kp~ and \vsys~are the same as for the low-radial-velocity-change case, with the transit duration of the injected signal multiplied to simulate the larger planet pixel shift of \npx = 25, in accordance with Figure \ref{fig:npx}, and consistent with the high-velocity case used in Figures \ref{fig:signal_degredation} and \ref{fig:signal_degredation_nin_nout_npx}. The number of principal components subtracted during detrending is 6, 9, and 7 for each of \ce{NH3}, \ce{CH4}, and \ce{H2O}, respectively. S/N values of 7.5, 5.7, and 4.3 are obtained for \ce{NH3}, \ce{CH4}, and \ce{H2O}, respectively, with the recovered \kp~constrained in each case.  It should be noted that the noise remaining in the data is greater for the high-velocity case than the low-velocity case, and therefore the recovered S/N is not proportional to the fraction of signal remaining, as given in Figure \ref{fig:signal_degredation}.}
    \label{fig:high_velocity}
\end{figure*}

\begin{figure*}[h]
    \centering
    \includegraphics[width=0.33\textwidth]{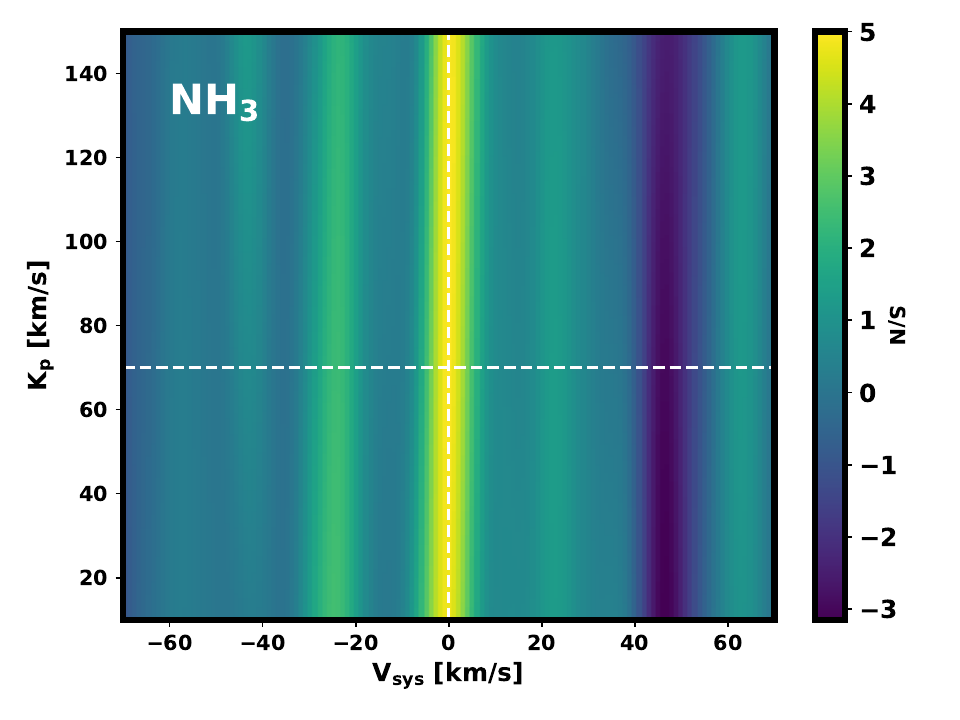}
    \includegraphics[width=0.33\textwidth]{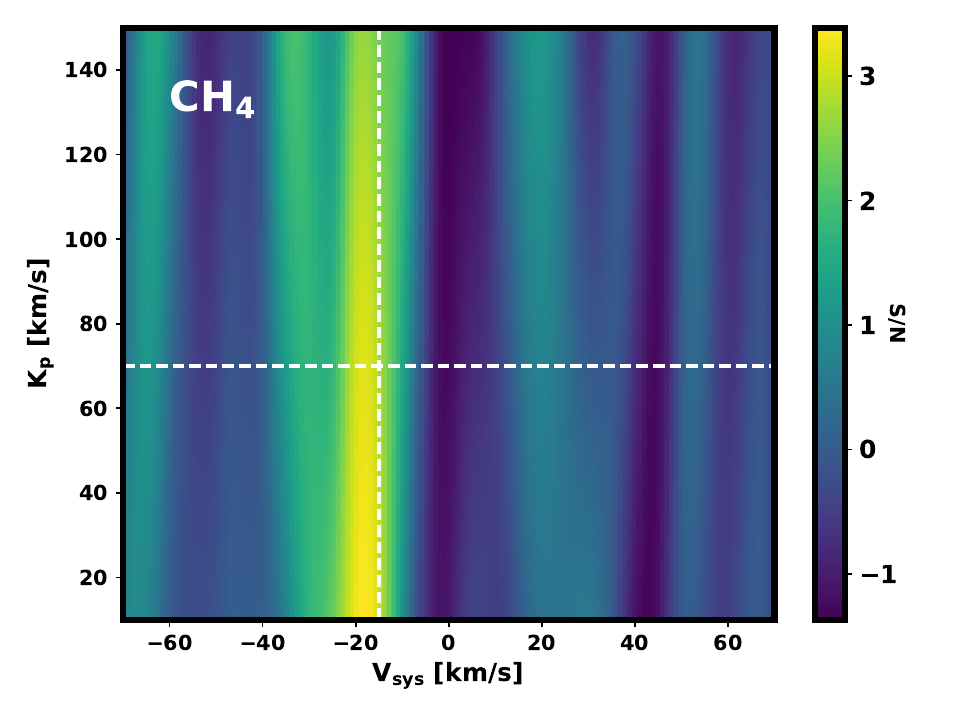}
    \includegraphics[width=0.33\textwidth]{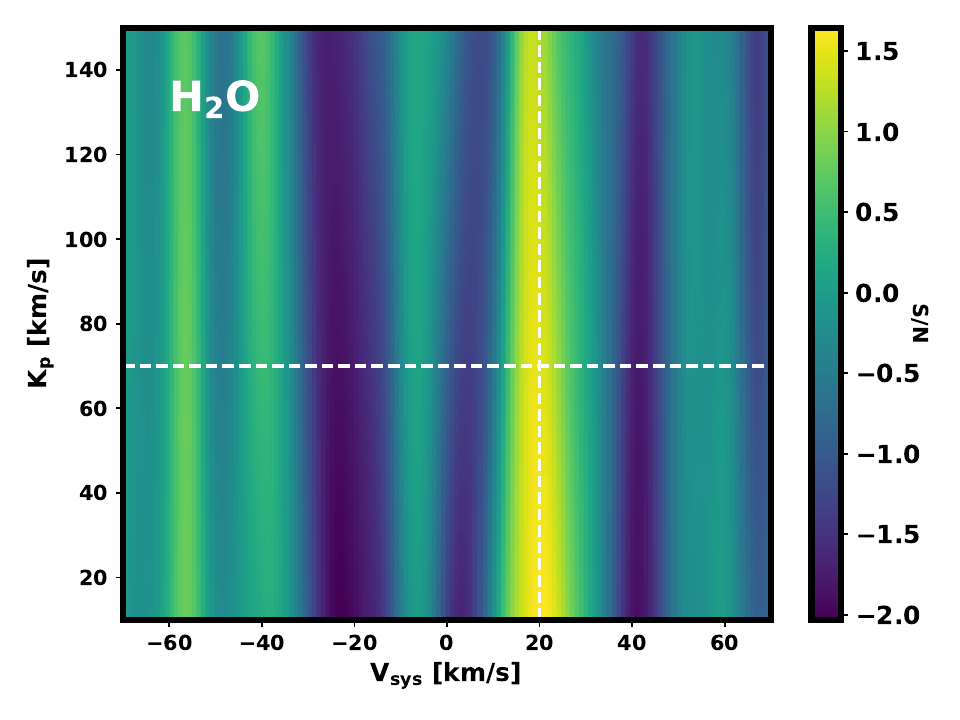}
    \caption{S/N maps obtained for each of \ce{NH3}, \ce{CH4}, and \ce{H2O}, when using the raw spectra rather than first dividing by the spectrum of an A0V telluric standard star, as was done throughout the rest of this work. The signals continue to be recovered, with S/N values consistent with before (5.0 for \ce{NH3}, 3.4 for \ce{CH4}, and 1.6 for \ce{H2O}).}
    \label{fig:raw}
\end{figure*}

Figure \ref{fig:high_velocity} shows the results for injection and recovery tests in the high-velocity regime, as described in Section \ref{low_velocity_regime}, for comparison with the low-velocity simulations in this work. Figure \ref{fig:raw} shows the results when raw spectra are used rather than the A0V-corrected spectra used throughout this work.

\end{document}